\documentclass[10pt,twocolumn]{article}
\usepackage{colonequals}
\usepackage{amsmath}
\usepackage{amssymb}
\usepackage{amsfonts}
\usepackage{supertabular}
\usepackage[T1]{fontenc}
\usepackage{txfonts}
\usepackage{latexsym}
\usepackage{color}
\usepackage[ruled,noend]{algorithm2e}
\LinesNumbered
\usepackage{url}
\usepackage{graphicx}
\usepackage{multirow}
\usepackage{microtype}
\usepackage{yfonts}
\usepackage{paralist}
\usepackage{afterpage}
\usepackage{balance}
\usepackage{array}
\SetKwFor{ForAll}{forall}{do}{end forall}
\usepackage[
draft,
bookmarks=false,
colorlinks=false,
pdftitle={Transient Reward Approximation for Continuous-Time Markov Chains},
pdfauthor={Ernst Moritz Hahn, Ralf Wimmer, Holger Hermanns, Bernd Becker},
pdfkeywords={Continuous-time Markov chains, 
  continuous-time Markov decision processes, 
  abstraction, 
  symbolic methods, 
  OBDDs
}
]{hyperref}
\usepackage[all]{hypcap}
\usepackage{stmaryrd}

\newenvironment{proofsketch}{\paragraph{Proof sketch:}}{\hfill$\square$}
\newenvironment{proof}{\paragraph{Proof:}}{\hfill$\square$}

\newtheorem{definition}{Definition}

\newtheorem{example}{Example}
\newtheorem{lemma}{Lemma}
\newtheorem{corollary}{Corollary}

\renewcommand{\gets}{\colonequals}

\newcommand{\fctcall}[1]{\ensuremath{\mathsf{#1}}}

\def\qedsymbol{\rule{2mm}{2mm}}
\def\qed{\hfill{\qedsymbol}\vskip 1pt}

\newcommand{\PRISM}{\textsc{Prism}\xspace}
\newcommand{\SIGREF}{\textsc{Sigref}\xspace}
\newcommand{\NUSMV}{\textsc{NuSMV}\xspace}

\DeclareMathOperator{\prob}{Pr}

\newcommand{\states} {{\ensuremath S}}
\newcommand{\state}  {{\ensuremath s}}

\newcommand{\partition}{\ensuremath{P}}

\newcommand{\distr}  {{\ensuremath {\mu}}}
\newcommand{\distrs} {{\ensuremath {\mathit{Distr}}}}

\newcommand{\nats}    {{\ensuremath {\mathbb{N}}}}
\newcommand{\reals}  {{\ensuremath {\mathbb{R}}}}

\newcommand{\sched}  {{\ensuremath {\sigma}}}

\newcommand{\timelimit}{\emph{--~Time out~--}}
\newcommand{\memlimit}{\emph{--~Memory out~--}}

\DeclareRobustCommand\sfrac[1]{\@ifnextchar/{\@sfrac{#1}}%
                                            {\@sfrac{#1}/}}
\def\@sfrac#1/#2{\leavevmode\kern.1em\raise.5ex
         \hbox{$\m@th\mbox{\fontsize\sf@size\z@
                           \selectfont#1}$}\kern-.1em
         /\kern-.15em\lower.25ex
          \hbox{$\m@th\mbox{\fontsize\sf@size\z@
                            \selectfont#2}$}}

\newcommand{\acts}{\mathit{Act}}
\newcommand{\pmat}{\mathbf{P}}
\newcommand{\rmat}{\mathbf{R}}
\newcommand{\dmodel}{\mathcal{D}}
\newcommand{\cmodel}{\mathcal{C}}
\newcommand{\diff}{\mathrm{d}}
\newcommand{\mpath}{\beta}
\newcommand{\urate}{\mathbf{u}}
\newcommand{\act}{\alpha}
\newcommand{\schedshr}{\Sigma_\mathit{HR}}
\newcommand{\schedscr}{\Sigma_\mathit{CR}}
\newcommand{\schedscd}{\Sigma_\mathit{CD}}
\newcommand{\schedhr}{{\sigma_\mathit{hr}}}
\newcommand{\schedcr}{{\sigma_\mathit{cr}}}

\newtheorem{proposition}{Proposition}
\newcommand{\expect}{\mathbf{E}}
\newcommand{\rew}{\mathbf{r}}
\newcommand{\frew}{\mathbf{r}_f}
\newcommand{\crew}{\mathbf{r}_c}
\newcommand{\irew}{\mathbf{r}_i}
\newcommand{\target}{\mathbf{B}}
\newcommand{\timeb}{\mathbf{t}}
\newcommand{\mvalue}{\mathbf{V}}
\newcommand{\precision}{\varepsilon}
\newcommand{\emb}{\mathrm{emb}}
\newcommand{\stopro}{X}
\newcommand{\ivall}{{\mathbf{I}^{\ell}}}
\newcommand{\ivalu}{{\mathbf{I}^u}}

\newcommand{\apart}{\mathfrak{P}}
\newcommand{\avar}{\mathfrak{x}}
\newcommand{\avars}{\mathfrak{V}}
\newcommand{\astates}{\mathfrak{A}}
\newcommand{\astate}{\mathfrak{z}}

\newcommand{\bdd}{\mathsf{b}}
\newcommand{\bddVars}{\mathsf{V}}
\newcommand{\bddVar}{\mathsf{x}}
\newcommand{\bddNodeInit}{\bddNode_{\mathrm{root}}}

\newcommand{\bddNodes}{\mathsf{N}}
\newcommand{\bddNode}{\mathsf{n}}
\newcommand{\bddHi}{\mathsf{h}}
\newcommand{\bddLo}{\mathsf{l}}
\newcommand{\bddNV}{\mathsf{v}}
\newcommand{\bddZero}{\mathsf{bdd}_{\mathsf{0}}}
\newcommand{\bddOne}{\mathsf{bdd}_{\mathsf{1}}}

\newcommand{\bddEval}{\mathsf{Val}}
\newcommand{\bddInit}{\mathsf{init}}
\newcommand{\bddSucc}{\mathsf{succ}}

\newcommand{\prismModel}{m}
\newcommand{\prismVar}{x}
\newcommand{\prismVars}{\mathit{Var}}
\newcommand{\prismInit}{\mathit{init}}
\newcommand{\prismSucc}{\mathit{succ}}
\newcommand{\prismNSucc}{\overline{\prismSucc}}
\newcommand{\prismCmds}{C}
\newcommand{\prismCmd}{c}
\newcommand{\prismCRew}{R_c}
\newcommand{\prismFRew}{R_f}

\newcommand{\samspace}[1]{\Omega_{#1}}

\newcommand{\refsec}[1]{\texorpdfstring{\hyperref[sec:#1]{Section~\ref*{sec:#1}}}{Section~\ref*{sec:#1}}}
\newcommand{\refdef}[1]{\texorpdfstring{\hyperref[def:#1]{Definition~\ref*{def:#1}}}{Definition \ref*{def:#1}}}
\newcommand{\reffig}[1]{\texorpdfstring{\hyperref[fig:#1]{Fig.~\ref*{fig:#1}}}{Fig.~\ref*{fig:#1}}}
\newcommand{\reflem}[1]{\texorpdfstring{\hyperref[lem:#1]{Lemma~\ref*{lem:#1}}}{Lemma~\ref*{lem:#1}}}
\newcommand{\refpro}[1]{\texorpdfstring{\hyperref[pro:#1]{Proposition~\ref*{pro:#1}}}{Proposition~\ref*{pro:#1}}}

\newcommand{\reftab}[1]{\texorpdfstring{\hyperref[tab:#1]{Table~\ref*{tab:#1}}}{Table~\ref*{tab:#1}}}
\newcommand{\refexa}[1]{\texorpdfstring{\hyperref[exa:#1]{Example~\ref*{exa:#1}}}{Example~\ref*{exa:#1}}}
\newcommand{\refeqn}[1]{\texorpdfstring{\hyperref[eqn:#1]{\eqref{eqn:#1}}}{\eqref{eqn:#1}}}
\newcommand{\refseqn}[1]{\texorpdfstring{\hyperref[eqn:#1]{Eqn.~\ref*{eqn:#1}}}{Eqn.~\ref*{eqn:#1}}}
\newcommand{\refscor}[1]{\texorpdfstring{\hyperref[cor:#1]{Cor.~\ref*{cor:#1}}}{Cor. \ref*{cor:#1}}}
\newcommand{\refalg}[1]{\texorpdfstring{\hyperref[alg:#1]{Algorithm~\ref*{alg:#1}}}{Algorithm~\ref*{alg:#1}}}
\newcommand{\refcor}[1]{\texorpdfstring{\hyperref[cor:#1]{Corollary~\ref*{cor:#1}}}{Corollary~\ref*{cor:#1}}}
\newcommand{\reflin}[1]{\texorpdfstring{\hyperref[lin:#1]{Line~\ref*{lin:#1}}}{Line~\ref*{lin:#1}}}

\newcommand{\defeq}{\,\stackrel{\mbox{\rm {\tiny def}}}{=}\,}

\newcommand{\partialto}{\rightharpoonup}

\newcommand{\dom}{\mathit{Dom}}
\newcommand{\sem}[1]{\ensuremath{\llbracket #1 \rrbracket}\xspace}
\newcommand{\tprob}{\pi}

\title{Transient Reward Approximation for\\ Continuous-Time Markov Chains}

\author{Ernst Moritz Hahn\\
  State Key Laboratory of Computer Science,\\
  Institute of Software, Chinese Academy of Sciences\\
  \and
  Holger Hermanns\\
  Saarland University\\
  Germany
  \and
  Ralf Wimmer\\
  Albert-Ludwigs-Universit\"at\\
  Freiburg, Germany
  \and
  Bernd Becker\\
  Albert-Ludwigs-Universit\"at\\
  Freiburg, Germany
}
\date{}

\begin{document}

\maketitle

\begin{abstract}
  \noindent We are interested in the analysis of very large continuous-time
  Markov chains (CTMCs) with many distinct rates. Such models arise
  naturally in the context of reliability analysis, e.\,g.,
  of computer network performability analysis, of power grids,
  of computer virus vulnerability, and in the study of
  crowd dynamics. We use abstraction techniques together with novel
  algorithms for the computation of bounds on the expected final and
  accumulated rewards in continuous-time Markov decision processes
  (CTMDPs). These ingredients are combined in a partly symbolic and
  partly explicit (symblicit) analysis approach. In particular, we
  circumvent the use of multi-terminal decision diagrams, because the
  latter do not work well if facing a large number of different
  rates. We demonstrate the practical applicability and efficiency of
  the approach on two case studies.
    \vspace{-1.6mm}
\end{abstract}

{\footnotesize
\paragraph*{\itshape Acknowledgements}
This work was partly supported by the German Research
Council (DFG) as part of the Transregional Collaborative
Research Centre ``Automatic Verification and Analysis of Complex
Systems'' (\href{www.avacs.org}{SFB/TR~14~AVACS}), the NWO-DFG \href{http://rocks.w3.rz.unibw-muenchen.de/}{bilateral project ROCKS},
\href{http://www.veriware.org/}{the ERC Advanced Grant VERIWARE},
the National Natural Science Foundation of China (NSFC) under grant No.\  61350110518 and No.\ 61450110461,
the CDZ project CAP (GZ 1023),
the Chinese Academy of Sciences Fellowship for International Young Scientists (Grant No.\ 2013Y1GB0006),
and has received funding from the European Union Seventh Framework Programme under 
grant agreement number 295261 as part of the
\href{http://www.meals-project.eu/}{MEALS project} and  under
grant agreement number 318490 as part of the
\href{http://www.sensation-project.eu/}{SENSATION project}.%

Part of this work was done while Ernst Moritz Hahn was with Saarland University and with the University of Oxford, United Kingdom

We thank Martin Neuh{\"a}u{\ss}er and Lijun Zhang for fruitful discussions
This article appeared in IEEE Transactions on Reliability, Volume 64, Issue 4.
}

\paragraph*{\itshape Abbreviations}\ \\
\begin{supertabular}{p{1.6cm}>{\raggedright\arraybackslash}p{5.7cm}}
ACTMC & abstract continuous-time Markov chain\\
BDD & binary decision diagram\\
CD & time-abstract, history-abstract, counting, deterministic scheduler\\
CR & time-abstract, history-abstract, counting, randomised scheduler\\
CSL & continuous stochastic logic\\
CTMC & continuous-time Markov chain\\
CTMDP & continuous-time Markov decision process\\
DTMC & discrete-time Markov chain\\
DTMDP & discrete-time Markov decision process\\
ECTMC & extended abstract continuous-time Markov chain\\
EVBDD & edge-valued decision diagram\\
HR & time-abstract, history-dependent, randomi\-sed scheduler\\
MDD & multiple-valued decision diagram\\
MTBDD & multi-terminal binary decision diagram\\
OBDD & reduced ordered binary decision diagram\\
PM & \PRISM model\\
symblicit & method combining symbolic and explicit aspects\\
ZDD & zero-suppressed decision diagram\\
\end{supertabular}

\bigskip

\paragraph*{\itshape Notation}\ \\
\begin{supertabular}{p{1.6cm}p{5.7cm}}
  $\distr$ & probability distribution\\
  $\distrs(A)$ & set of probability distributions over the set $A$\\
  $\dmodel$ & discrete-time Markov model\\
  $\states$ & states of a Markov model\\
  $\pmat$ & probability matrix of a Markov model\\
  $\stopro^{\dmodel,\state_0}$ & stochastic process of a Markov model\\
  $\state$ & state of a Markov model\\
  $(\Omega,\Sigma)$ & measurable space\\
  $\prob$ & probability measure\\
  $\expect$ & expectation\\
  $\rmat$ & rate matrix of a Markov model\\
  $\urate$ & uniformisation rate\\
  $\acts$ & action set of a Markov decision process\\
  $\act$ & action of a Markov decision process\\
  $\widehat{\acts}$ & action set of an extended abstract Markov chain\\
  $\ivall, \ivalu$ & intervals of an extended abstract Markov chain\\
  $\hat{\act}$ & action of an extended abstract Markov chain\\
  $\sem{\cmodel}$ & continuous-time Markov decision process semantics of an extended abstract Markov chain\\
  $\sched$ & scheduler of a Markov decision process\\
  $\mpath$ & history of a Markov model\\
  $\schedshr$ & set of time-abstract, history-dependent, randomised schedulers\\
  $\schedscr$ & set of time-abstract, history-abstract, counting, randomised schedulers\\
  $\schedscd$ & set of time-abstract, history-abstract, counting, deterministic schedulers\\
  $\sched$ & scheduler\\
  $\rew = (\crew, \frew)$ & reward structure with cumulative reward rate $\crew$, and final reward value $\frew$\\
  $\frew^{\max}$ & maximal final reward value over model states\\
  $\crew^{\max}$ & maximal cumulative reward rate over model states\\
  $\timeb$ & time bound\\
  $\mvalue$ & reward value of a stochastic process or Markov model\\
  $\prismModel$ & \PRISM model\\
  $\prismVars$ & variables of a \PRISM model\\
  $\prismInit$ & initial state of a \PRISM model\\
  $\prismCmds$ & commands of a \PRISM model\\
  $\prismSucc$ & successors function of a \PRISM model\\
  $\prismCRew$ & cumulative rewards of a \PRISM model\\
  $\prismFRew$ & instantaneous rewards of a \PRISM model\\
  $\prismNSucc$ & qualitative successor function of a \PRISM model\\
  $\states_{\prismModel}$ & reachable states of a \PRISM model\\
  $\cmodel_\prismModel$ & induced CTMC of a \PRISM model\\
  $\rew_\prismModel$ & induced reward structure of a \PRISM model\\
  $\apart$ & partitioning of state space of a \PRISM model\\
  $\astate$ & abstract state\\
  $\bddVars$ & BDD variables\\
  $\bddVar$ & BDD variable\\
  $\bdd$ & graph of BDD\\
  $\bddNodes$ & set of nodes of a BDD\\
  $\bddNodeInit$ & root node of a BDD\\
  $\bddNV(\bddNode)$ & label of a BDD terminal node $\bddNode$\\
  $\bddHi(\bddNode)$ & high successor of a BDD node $\bddNode$\\
  $\bddLo(\bddNode)$ & low successor of a BDD node $\bddNode$\\
  $v$ & variable valuation of a BDD\\
  $\bddEval$ & set of variable valuations of a given BDD\\
  $\sem{\bdd}$ & function represented by a BDD $\bdd$\\
  $\bddZero, \bddOne$ & zero, and one BDDs\\
  $\bdd_\apart$ & BDD representation of a state space partitioning\\
  $\avars$ & set of BDD variables to encode the number of an abstract state\\
  $\avar$ & BDD variable to encode the number of an abstract state\\
  $\phi_\lambda$ & Poisson distribution with rate $\lambda$\\
  $\psi_\lambda$ & cumulative Poisson distribution with rate $\lambda$\\
  $\precision$ & precision to compute value bounds\\
  $\dom$ & domain of a partial function\\
\end{supertabular}

\section{Introduction}
\label{sec:introduction}
The analysis of large Markov chains is a
recurring challenge in many important areas ranging from computer
network dependability and performance~\cite{sand87a,haverkort-book} to
quantitative security~\cite{KopfB11}. To evaluate properties of such
systems, a standard approach is to perform numerical analysis,
nowadays often embedded in a stochastic model
checker~\cite{KNP11,KatoenZHHJ11,CiardoMW09,DeavoursCCDDDSW02}. At its
core, the model checker has to operate with a very large matrix
induced by the Markov chain. In this context, the use of symbolic
representations, in particular variations of decision diagrams, such
as multi-terminal decision
diagrams (MTBDDs)~\cite{Parker02,HermannsKNPS03}, multiple-valued decision diagrams (MDDs)~\cite{WanCM11}, or
zero-suppressed decision diagrams (ZDDs)~\cite{LampkaSOB10}, have made it possible to store and manipulate
very large matrices in a symbolic manner (either of
transition rates or just of adjacency). Many of the applications
occurring in practice lead to very large continuous-time Markov chains
(CTMCs) that nevertheless contain only a very small number of
different transition rates. This is a primary reason why decision

diagrams, where distinct rates are stored as distinct values in the
structure, are effective. Whenever there are many pairwise different
rates occurring, the decision diagram degenerates to a decision tree,
and thus its size explodes. Edge-valued binary decision
diagrams (EVBDDs)~\cite{LaiPV96} often can avoid this representation
explosion, at the price of a more involved reconstruction of matrix
entries. This trade-off makes them less suited for direct numerical
computations, as needed for the model checking of CTMCs. Therefore,
models with a large number of different rates are a
notorious problem for symbolic representations, and hence for the
stochastic model checkers available to date.

However, there is a growing spectrum of important applications that
give rise to excessive numbers of distinct rates.
Computer network performability analysis \cite{HaverkortHK00,ClothH05,BaierHHHK12,DBLP:journals/sqj/GokhaleLT04,DBLP:books/daglib/0079988},
power grid stability~\cite{DBLP:conf/nca/Sanders12,wsc12},
crowd dynamics~\cite{MassinkLBHH12,MassinkLBH11},
as well as (computer) virus epidemiology~\cite{XuLZ12,ZEB+11,YA11}
are important examples where Markov models are huge, and rates change
from state to state. The study of these phenomena is of growing
importance for the assurance of their reliability. Several
of these examples can in some way be regarded as Markov population
models~\cite{HenzingerJW11,HillstonTG12},
where the rates change with population counts, similar to models
appearing in systems biology \cite{MateescuWDH10}, and also in classical performance and
dependability engineering \cite{HaverkortHK00,ClothH05}.

This paper targets the analysis of transient properties of CTMCs with
both a large number of states as well as a large number of distinct
transition rates. It presents a combination of abstraction techniques,
an explicit representation of a small abstract model, and symbolic
techniques. The latter use reduced ordered binary decision diagrams (OBDDs), not MTBDDs or
MDDs. As our method involves both symbolic and explicit state space representations,
we call it \emph{symblicit}.
The abstraction method relies on
visiting all concrete states of the abstract model to obtain bounds on
the transition matrix, but without having to store the state space
explicitly. We also present ideas how to speed up this admittedly
time-consuming process. On the one hand, the approach can be seen
as a continuation of our previous work on symblicit
algorithms~\cite{WimmerBBHCHDT10,CrouzenHHDTWBB11}. On the other hand,
we harvest work done on the abstraction of Markov chains to abstract
Markov chains or Markov decision
processes~\cite{Klink10,Smith10,KatoenKLW07,Buchholz11,AlfaroR07,KattenbeltKNP10,HermannsWZ08}.

A number of related methods exist.
Our solution method uses results for explicit-state model analysis using \emph{uniformisation (randomisation)}~\cite{Jensen53,GrossM84}.
In this method, a continuous-time Markov model is transformed into a discrete-time Markov model and a Poisson process.
Intuitively, the discrete-time model describes in which state the model resides after a state change, while the Poisson process describes the process of the state changes.
Analyses using uniformisation then usually perform computations to obtain intermediate results on the discrete-time model, and later combine these results by weighting them using the Poisson process.
As the number of possible state changes is unbounded, this process needs to be truncated, thereby only considering a finite number of potential state changes. However, the probability residing beyond that truncation point can be bounded~\cite{FoxG88}, so that the precision of the results obtained via uniformisation can be precomputed.
Uniformisation is well understood, numerically stable, and generally performs well in practice for non-stiff models.
Because of this, a number of contributions have since been based on this method.
Advanced methods for explicit-state Markov reward models have been pioneered by Trivedi et al.~\cite{TrivediRS87}. Based on a variant of uniformisation using quasi-stationary detection, Carrasco has developed a method \cite{Carrasco04} to speed up the
computation of transient reward properties for large stiff models where the state space can be partitioned into a transient set and an absorbing state set. Models with a similar structure are amenable to the efficient analysis of cumulative reward properties with a very general state-based reward
notation~\cite{CarrascoS11}.

We also build on many ideas for analyses using abstract Markov chains
by Klink et al.~\cite{Klink10,KatoenKLW07}. This paper extends their
works by describing a widely applicable abstraction method, and also
by handling more general transient properties.

MTBDD-based methods~\cite{Parker02,HermannsKNPS03} work well for some
models, but have the disadvantages described above.
The method of Wan et al.~\cite{WanCM11} uses a slightly different data
structure to represent concrete models, and focuses on steady-state
properties rather than transient ones. It does not rely on Markov
decision models, and, though it works well in practice for certain
model classes, it cannot guarantee safe bounds for properties of the
concrete model.

There are techniques in which a symbolic representation of the transition
matrix is used, but in which values assigned to the states of the model
(such as probabilities) have to be stored explicitly and separately for
each state of the model. Examples include the so-called hybrid method
\cite{Parker02,KwiatkowskaNP04}, other variants of decision diagrams
\cite{LampkaSOB10}, and also methods using Kronecker representations
\cite{Dayar12,BuchholzCDK00,BenoitPS06,FernandesPS98,BuchholzK04}.
These kinds of methods are more precise, and might be faster than the
method we propose. They are however not applicable in case the state space
is excessively large, too large to store one value per state.

Smith~\cite{Smith10} developed means for the compositional abstraction
of CTMCs given in a process calculus. This way, he obtains an
abstract Markov chain, which is then analysed by a method of Baier et
al.~\cite{BaierHKH05} to obtain bounds for the time-bounded
reachability probability. Our approach uses a different abstraction
method, and can handle a more general class of properties.

A paper by Buchholz~\cite{Buchholz11} describes how bounds on long-run
average (thus, non-transient) properties can be obtained from abstract
Markov chains. It is based on a combination of policy and value
iteration~\cite{Puterman94}, and discusses the applicability of
several variants of these methods on typical examples from queueing
theory and performance evaluation.

The magnifying-lens abstraction~\cite{AlfaroR07} by de Alfaro et
al.~is similar to our approach in that it also builds on (repeated)
visits of concrete model states without storing the whole concrete
state space. It discusses a different model, discrete-time Markov
decision processes (DTMDPs), and a different property, time-unbounded
reachability probabilities.

D'Argenio et al.~\cite{DArgenioJJL01} discuss how DTMDPs given as
MTBDDs can be abstracted to obtain a smaller abstract model, which is
also a DTMDP, but small enough to be represented explicitly. In
addition, a heuristic abstraction refinement method is presented. The
target there was to obtain bounds for unbounded reachability
probabilities. Works by Hermanns et al.~\cite{HermannsWZ08} and by
Kattenbelt et al.~\cite{KattenbeltKNP10} later developed methods to
use probabilistic games to provide tighter value bounds and predicate
abstraction to handle larger or even infinitely large models, as well
as refinement methods based on these frameworks. In contrast to the
state of the art for discrete-time models, the discussed refinement
method we consider is more preliminary.

Other methods work with a finite subset of concrete states of the
model under consideration, rather than subsuming concrete states in
abstract ones. There exists a wide range of methods based on this principle
for the analysis of CTMCs \cite{Grassmann91,MoorselS94,MunskyK06,HenzingerMW09}.
Recently, this approach was extended to infinite-state Markov decision processes~\cite{Buchholz12}.
Here, two finite submodels are constructed which guarantee to bound the
values over all policies from below and above. They can also be used
to obtain a policy which is $\varepsilon$-optimal in the original model.

Such methods are applicable if during a transient analysis the
probability mass stays concentrated on a small subset of states at
each point of time.  If, however, the probability spreads evenly among
too many states of the model, then such methods are not appropriate. The
reason is that, in this case, either too many states need to be stored, 
or a too large amount of the probability mass is lost as too many
states have to be disregarded.

In \refsec{preliminaries}, we
provide basic notations, and describe the symbolic data structures used
for the later abstraction. We also describe the formal models we use,
as well as the properties we are interested in. \refsec{algorithms}
describes algorithms to efficiently obtain an abstract model from a
description of a concrete model, and discusses how they can be used to
bound properties of the concrete model. In \refsec{case-studies}, we
apply this method on two case studies from the area of computer network
performability analysis, thus to show its practical applicability. Finally,
\refsec{conclusion} concludes the paper.

\section{Preliminaries}
\label{sec:preliminaries}
\noindent This section introduces basic notations, and formally defines the
models and data structures that are used in the later parts of the
paper.

In Subsection~\ref{subsec:stochastic_models}, we discuss the stochastic models that build the theoretical foundation of the method described in this paper.
We start by describing Markov chains, the mechanism in which the models to analyse are formulated originally (Definitions~\ref{def:dtmc} and \ref{def:ctmc}).
Next, we define Markov decision processes, which extend Markov chains by nondeterministic decisions (Definitions~\ref{def:dtmdp} through \ref{def:emb}).
We then state a model type which allows a compact representation of Markov decision processes with an infinite number of nondeterministic choices per
state, and which we will later use to obtain abstractions of Markov chains (Definition~\ref{def:imc}).
Afterwards, we discuss how the nondeterminism of Markov decision processes can be resolved using an entity called scheduler (Definitions~\ref{def:sched-hrs} through \ref{def:induced-ctmc}). 
This process is necessary to obtain a stochastic process, which is needed to reason about their properties.
Finally, we assign rewards to the states of the discussed stochastic models (Definition~\ref{def:reward-structure}).
These reward structures allow us to define time-dependent reward values (Definition~\ref{def:value}), which allow us to express a wide variety of interesting properties.

Subsection~\ref{subsec:prism_gcl} discusses a high-level specification language (Definition~\ref{def:prism-model}), the semantics of which is again a stochastic model (Definition~\ref{def:induced-ctmc-rewards}).
Such a language allows us to express stochastic models in a compact way, and is thus both more memory-efficient and easier to read by humans than an explicit state-wise representation.
However, in contrast to an explicit-state representation (or certain symbolic representations), it is not amenable to a direct analysis of its properties.
In this paper, we target to avoid constructing the semantics of such high-level models explicitly.
Instead, we define abstract state spaces (Definition~\ref{def:partitioning}) in which we subsume sets of concrete states to abstract states.
We emphasise that, with the method of this paper, we do not have to explicitly store all such concrete states at the same time to generate abstractions.

In Subsection~\ref{subsec:obdds}, we describe the concrete data structure we are going to use to store such a partitioning in a compact way (Definition~\ref{def:bdd-partitioning}).
In addition, we use this data structure to store whether there is a non-zero rate between two states of the semantics of a high-level model (Definition~\ref{def:bdd-repr}).
For our method, we do not need to store concrete values of non-zero rates in this representation, however.
Thus, we can use simple OBDDs (Definitions~\ref{def:bdd} and \ref{def:reduced_ordered}), because this data structure already fulfils these requirements.

\subsection{Stochastic Models}
\label{subsec:stochastic_models}

\noindent A \emph{distribution} over a finite or countable set $A$ is a function $\distr\colon A \to
[0,1]$ such that $\sum_{a \in A} \distr(a) = 1$. By $\distrs(A)$, we
denote the set of all distributions over $A$. 

The simplest stochastic model we consider is as follows.
\begin{definition}
  \label{def:dtmc}
  A \emph{discrete-time Markov chain (DTMC)} is a tuple $\dmodel = (\states, \pmat)$ where
  \begin{itemize}
  \item $\states$ is a finite set of \emph{states}, and
  \item $\pmat\colon (\states \times \states) \to [0,1]$ is the \emph{probability matrix} such that $\sum_{\state' \in \states} \pmat(\state, \state') = 1$ for all $\state \in \states$.
  \end{itemize}
  By $\stopro^{\dmodel,\state_0}\colon (\samspace{\dmodel} \times \nats) \to \states$ with $\state_0 \in \states$ we denote the unique stochastic process~\cite{Stewart94} of $\dmodel$ with initial state $\state_0$, where $\samspace{\dmodel}$ is the sample space to be used.
\end{definition}
The time in a DTMC proceeds in discrete steps, and in each step a transition with non-zero probability is taken.
At step $0$ the model starts in a given initial state $\state_0 \in \states$.
The model moves to the next state, and will be in $\state_1$ with probability $\pmat(\state_0, \state_1)$ for all $\state_1 \in \states$.
From there, again the next state is chosen according to $\pmat$, and so on.

By $\prob$, we denote the \emph{probability measure} on the measurable spaces $(\Omega_{\dmodel},\Sigma_{\dmodel})$ 
of the DTMC $\dmodel$ under consideration, with sample space $\Omega_{\dmodel}$, and set $\Sigma_{\dmodel}$ of events,
which is defined by the standard cylinder set construction over finite paths~\cite{KSK66}.
For instance, $\prob(\stopro^{\dmodel,\state_0}_n = \state_1 \vee \stopro^{\dmodel,\state_0}_{n+1} = \state_2)$
describes the probability that, having started in state $\state_0$, in step $n$ we are in $\state_1$, or in step $n+1$ we are in $\state_2$.
For a measurable function $X\colon \Omega_{\dmodel} \to \reals$, we thus also have an \emph{expectation} $\expect(X) \defeq \int_{\Omega_{\dmodel}} X(\omega) \prob(\diff \omega)$.
For instance, consider $X \defeq \sum_{i=0}^{n-1} (f \circ \stopro^{\dmodel,\state_0}_i)$ such that $f(\state_1) \defeq 1$, and $f(\state) \defeq 0$ for $\state \neq \state_1$.
Then $\expect(X)$ denotes the average number of steps within the first $n$ steps in which the DTMC is in $\state_1$, under the condition that we started in $\state_0$.

We now discuss our basic stochastic model described informally in the introduction.
\begin{definition}
  \label{def:ctmc}
  A \emph{(uniform) continuous-time Markov chain (CTMC)} is a tuple $\cmodel = (\states, \rmat)$ where 
  \begin{itemize}
  \item $\states$ is a finite set of \emph{states}, and
  \item $\rmat\colon (\states \times \states) \to \reals_{\geq 0}$ is the
    \emph{rate matrix} such that there is a \emph{uniformisation rate} $\urate(\cmodel) > 0$
    with $\sum_{\state' \in \states}
    \rmat(\state, \state') = \urate(\cmodel)$ for all $\state \in \states$.
  \end{itemize}
\end{definition}
If $\cmodel$ is clear from context, we write $\urate$ instead of $\urate(\cmodel)$.
In a \emph{non-uniform CTMC}, the requirement $\sum_{\state' \in \states} \rmat(\state, \state') = \urate(\cmodel)$ does not hold for all states.
Every finite non-uniform CTMC can be transformed into an equivalent uniform CTMC with the same stochastic behaviour by increasing 
$\rmat(\state, \state)$ such that the total sum is the same for all states~\cite{Stewart94}.
We require uniformity only for ease of presentation; it does not restrict the applicability of the methods developed here to general CTMCs.

The behaviour of a CTMC $\cmodel = (\states, \rmat)$ is similar to a DTMC.
However, the durations until state changes are now real numbers.
They are chosen according to statistically independent negative exponential distributions with parameter $\urate$.
Thus, the probability that a state change takes place within time $t$ is $1 - \mathrm{e}^{-\urate\cdot t}$.
The successor state is then selected according to the distribution $\distr\colon \states \to [0,1]$ with $\distr(\state') = \rmat(\state_0,\state')/\urate$ for all $\state' \in \states$.
We assume that the process runs until a certain point of time $\timeb$ is reached.
By $\stopro^{\cmodel,\state_0}\colon (\samspace{\cmodel} \times \reals_{\geq 0}) \to \states$ with $\state_0 \in \states$,
we denote the uniquely defined stochastic process~\cite{Stewart94} of $\cmodel$ with initial state $\state_0$,
where $\samspace{\cmodel}$ is the sample space to be used.
As for DTMCs, we assume that we have probability measures and expectations on the sample spaces.
\begin{figure*}
\centering
\includegraphics{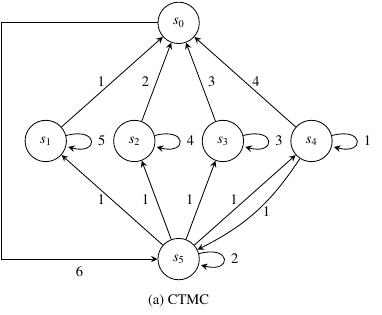}
\hfill
\includegraphics{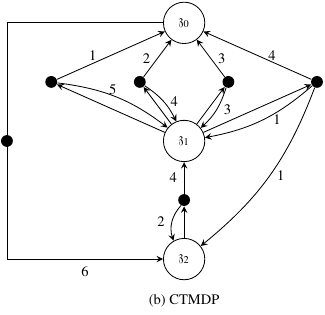}
\hfill
\includegraphics{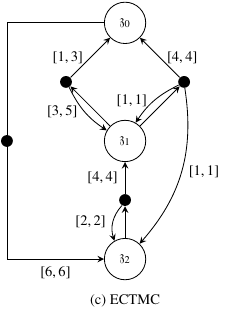}
\ \\\ \\
\includegraphics{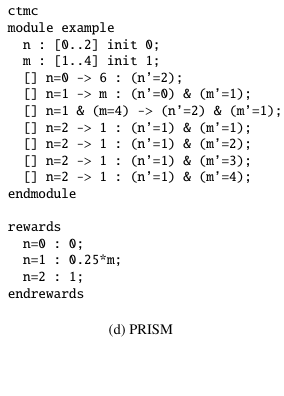}
\hfill
\includegraphics{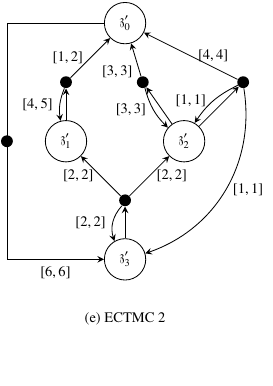}
\hfill
\includegraphics{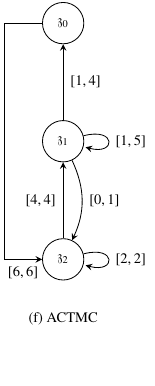}
\hfill
\includegraphics{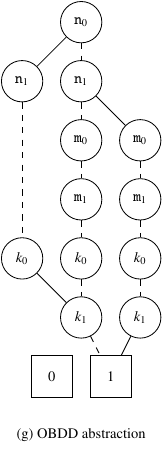}
\caption{\label{fig:examples}Example models.}
\end{figure*}

\begin{example}
  \label{exa:ctmc}
  In \reffig{examples}(a), we give an example for a CTMC.
  We represent its states as circles, and non-zero rates between states are given as arrows labelled with the rates.
  The uniformisation rate is $6$.
\end{example}

We need to specify another discrete- and a continuous-time model~\cite{Howard60,Bert05}, which will later be used to abstract large CTMCs.
In addition to stochastic behaviour, these models also feature a \emph{nondeterministic} choice over the successor distributions.
Nondeterministic choices are choices which cannot be assigned a probability a priori.
Instead, different stochastic behaviours result according to the resolution of the nondeterminism.
\begin{definition}
  \label{def:dtmdp}
  A \emph{discrete-time Markov decision process (DTMDP)} is a tuple $\dmodel = (\states, \acts, \pmat)$ where $\states$ is as in \refdef{dtmc},
  \begin{itemize}
  \item $\acts$ is a set of \emph{actions}, and
  \item $\pmat\colon (\states \times \acts \times \states) \to [0,1]$ is the \emph{probability matrix} such that $\sum_{\state' \in \states} \pmat(\state,\act,\state') \in \{0, 1\}$ for all $\state \in \states$ and $\act \in \acts$. 
  \end{itemize}
  For $\state \in \states$, we denote the set of \emph{enabled} actions with $\acts(\state) \defeq \bigl\{ \act \in \acts\,\big|\,\sum_{\state' \in \states}  \pmat(\state,\act,\state') = 1 \bigr\}$.
  We require that either $|\acts|<\infty$, or that for all $\state \in \states$ and all $p\colon \states \to \reals_{\geq 0}$ the set $\bigl\{\sum_{\state' \in \states} \pmat(\state, \act, \state') \cdot p(\state')\,\big|\,\act \in \acts(\state) \bigr\}$ is compact.
\end{definition}
The behaviour of a DTMDP is such that, upon entering a state $\state \in \states$, an action $\act \in \acts(\state)$, or possibly a distribution over actions, is chosen.
This choice determines the probabilities of the state the model moves to in the next time step.
Notice that we do indeed allow uncountably many actions, with the given restriction.

\begin{definition}
  \label{def:ctmdp}
  A \emph{(uniform) continuous-time Markov decision process (CTMDP)} is a tuple $\cmodel = (\states, \acts, \rmat)$.
  Here, $\states$ and $\acts$ are as in \refdef{dtmdp}.
  By $\rmat\colon (\states \times \acts \times \states) \to \reals_{\geq 0}$, we denote
  the \emph{rate matrix} such that there is a fixed value $\urate(\cmodel)$ with
  $\sum_{\state' \in \states} \rmat(\state,\act,\state') \in \bigl\{0, \urate(\cmodel)\bigr\}$ for all $\state \in \states$ and $\act \in \acts$.
  If $\cmodel$ is clear from the context, we write $\urate$ instead of $\urate(\cmodel)$.
  For $\state \in \states$, we denote the set of \emph{enabled} actions with $\acts(\state) \defeq \bigl\{ \act \in \acts \,\big|\,\sum_{\state' \in \states} \rmat(\state,\act,\state') = \urate \bigr\}$.
  We require that either $|\acts|<\infty$, or that for all $\state \in \states$ and all $p\colon \states \to \reals_{\geq 0}$ the set $\bigl\{\sum_{\state' \in \states} \rmat(\state, \act, \state')/\urate \cdot p(\state') \,\big|\,\act \in \acts(\state) \bigr\}$ is compact.
\end{definition}
As in a DTMDP, upon entering a state $\state$, an action $\act \in \acts(\state)$ (or a distribution over this set) is chosen to determine the distribution over the successor states.
As for CTMCs, the model moves to this successor state after a time given according to the negative exponential distribution with parameter $\urate$.

\begin{example}
  \label{exa:ctmdp}
  In \reffig{examples}(b), we give an example for a CTMDP.
  The nondeterministic choices, available in each state, are represented 
  by the arrows leading to a filled circle. From such a circle, a 
  distribution leads to the successor states. As for the CTMC in \refexa{ctmc}, 
  its uniformisation rate is $6$.
\end{example}

We need the following transformation from continuous-time to discrete-time models.
\begin{definition}
  \label{def:emb}
  Given a CTMDP $\cmodel = (\states, \acts, \rmat)$, the \emph{embedded DTMDP} is defined as
  $\emb(\cmodel) \defeq (\states, \acts, \pmat)$ with $\pmat(\state, \act, \state') \defeq \rmat(\state, \act, \state')/\urate$
  for all $\state, \state \in \states$, and $\act \in \acts(\state)$.
\end{definition}

We introduce a formalism to specify CTMDPs, extending the \emph{abstract Markov chains} by Klink et al.~\cite{Klink10,KatoenKLW07,KozineU02}, which is also a specific form of a \emph{constraint Markov chain}~\cite{CaillaudDLLPW10}.
The purpose of this model is to efficiently represent CTMDPs with a large number of actions.
Instead of explicitly enumerating all possible choices over successor distributions, it allows us to specify lower and upper bounds on the rates between states.
\begin{definition}
  \label{def:imc}
  An \emph{extended abstract continuous-time Markov chain (ECTMC)} is a tuple 
  $\cmodel = (\states, \widehat{\acts}, \ivall, \ivalu)$ where $\states$ is 
  as in \refdef{dtmdp}, and $\widehat{\acts}$ is a finite set of actions.
  We consider the uniformisation rate $\urate(\cmodel)$ of the model.
  The \emph{intervals} are partial functions of the form 
  $\ivall, \ivalu\colon (\states \times \widehat{\acts}) \partialto (\states \to [0,\urate])$.
  We require $\ivall$ and $\ivalu$ to have the same domain.
  In addition, for each $\state \in \states$, there must be at least 
  one $\hat{\act}\in\widehat{\acts}$ such that $\ivall(\state, \hat{\act})$ is defined.

  The \emph{CTMDP semantics} of an ECTMC is defined as $\sem{\cmodel} \defeq (\states, \acts, \rmat)$.
  We have $(\hat{\act},v) \in \acts$ if $\hat{\act} \in \widehat{\acts}$, and if $v$ is of the form $v\colon \states \to [0,\urate]$ with $\sum_{\state \in \states} v(\state) = \urate$.
  It is $(\hat{\act}, v) \in \acts(\state)$ if $\ivall(\state, \hat{\act})$ and $\ivalu(\state,\hat{\act})$ are defined,
  and $v(\state') \in \bigl[\ivall(\state, \hat{\act})(\state'), \ivalu(\state, \hat{\act})(\state')\bigr]$ for all $\state' \in \states$.
  We let $\rmat\bigl(\state, (\hat{\act}, v), \state'\bigr) \defeq v(\state')$.
\end{definition}
An ECTMC thus represents a CTMDP in which, for each state $\state$, one chooses a possible action $\hat{\act}$ of $\widehat{\acts}$.
In addition, one has to choose an assignment of successor rates which fulfill the requirement on the intervals.
This way, the action set is uncountably large, but satisfies the requirements of \refdef{ctmdp}.
The difference to the model of Klink et al.~is the choice of $\hat{\act} \in \widehat{\acts}$ before the choice of the successor rates.
This difference allows us to obtain more precise abstractions than we could obtain if we were using (non-extended) abstract Markov chains, while 
it still allows us to implement efficient analysis methods, as seen later in Sections~\ref{sec:algorithms} and \ref{sec:case-studies}.
\begin{example}
  \label{exa:ectmc}
  We give an example for an ECTMC in \reffig{examples}(c) with uniformisation rate $6$. Compared to the
  CTMDP in \reffig{examples}(b), the ECTMC allows rate intervals instead of rates.
\end{example}

To obtain a stochastic process from nondeterministic models, the nondeterminism must be resolved.
\emph{Schedulers} (or \emph{policies}) formalise the mechanism to do so.
Below, we define the most powerful class of schedulers we consider in this paper, and the stochastic processes they induce.
A scheduler of this class can resolve the nondeterminism according to the states and actions (and their sequence) that were visited before the model moved to the current state.
It may also decide not to pick one specific action, but rather involve a probabilistic choice over the enabled actions of a state.
It is however neither aware of the exact time at which former events happened nor of the current time.
\begin{definition}
  \label{def:sched-hrs}
  A \emph{time-abstract, history-dependent, ran\-dom\-ised scheduler (HR)} for a DTMDP
  $\dmodel = (\states, \acts, \pmat)$ or a CTMDP $\cmodel = (\states, \acts, \rmat)$
  is a function $\sched\colon \bigl((\states \times \acts)^* \times \states\bigr) \to \distrs(\acts)$
  such that, for all $\mpath \in (\states \times \acts)^*$, and $\state \in \states$,
  we have that if $\sched(\mpath, \state)(\act) > 0$ then $\act \in \acts(\state)$.
  With  $\schedshr$, we denote the set of all HRs.
\end{definition}

\begin{definition}
  Assume we are given a CTMDP $\cmodel = (\states, \acts, \rmat)$, and a HR
  $\sched\colon \bigl((\states \times \acts)^* \times \states\bigr) \to \distrs(\acts)$. We
  define the \emph{induced CTMC} as $\cmodel_\sched \defeq (\states', \rmat')$
  with
  \begin{itemize}
  \item $\states' \defeq (\states \times \acts)^* \times \states$,
  \item $\rmat'\bigl((\mpath,\state), (\mpath, \state, \act,\state')\bigr) \defeq
    \sched(\mpath, \state)(\act) \cdot \rmat(\state, \act, \state')$ for $\mpath
    \in (\states \times \acts)^*$, $\state, \state' \in \states$, $\act \in
    \acts$, and $\rmat'(\cdot,\cdot) \defeq 0$ otherwise.
  \end{itemize}
  Let $\stopro^{\cmodel_\sched,\state_0}\colon (\samspace{\cmodel_\sched} \times
  \reals_{\geq 0}) \to \bigl((\states \times \acts)^* \times \states\bigr)$ be the stochastic
  process of the CTMC $\cmodel_\sched$ with initial state $\state_0 \in \states$,
  and let $f\colon \bigl((\states \times \acts)^* \times \states\bigr) \to \states$ with
  $f(\mpath, \state) \defeq \state$.
  The \emph{induced stochastic process} $\stopro^{\cmodel,\sched,\state_0}\colon
  (\samspace{\cmodel_\sched} \times \reals_{\geq 0}) \to \states$ of $\cmodel$ and
  $\sched$ starting in $\state_0$ is then defined as
  $\stopro^{\cmodel,\sched,\state_0}_t \defeq f \circ
  \stopro^{\cmodel_\sched,\state_0}_t$ for $t \in \reals_{\geq 0}$.
  Definitions for DTMDPs are likewise using $\pmat$ instead of $\rmat$.
\end{definition}

We specify a simpler subclass of the schedulers of \refdef{sched-hrs}.
Schedulers of this class are only aware of the number of state changes that have
happened so far, and may only choose a specific successor distribution rather
than a distribution over them.
\begin{definition}
  \label{def:sched-cd}
  A \emph{time-abstract, history-abstract, counting, deterministic scheduler
  (CD)} for a DTMDP $\dmodel = (\states, \acts, \pmat)$ or a CTMDP $\cmodel =
  (\states, \acts, \rmat)$ is a function $\sched\colon (\states \times \nats) \to
  \acts$ such that for all $\state \in \states$ and $n \in \nats$ if
  $\sched(\state, n) = \act$ then $\act \in \acts(\state)$.
  With  $\schedscd$ we denote the set of all CDs.
\end{definition}

\begin{definition}
  \label{def:induced-ctmc}
  Assume we are given a CTMDP $\cmodel = (\states, \acts, \rmat)$, and a CD $\sched\colon (\states\times \nats) \to \acts$.
  We define the \emph{induced CTMC} as $\cmodel_\sched \defeq (\states', \rmat')$ with
  \begin{itemize}
  \item $\states' \defeq \states \times \nats$,
  \item $\rmat'\bigl((\state, n), (\state', n+1)\bigr) \defeq \rmat\bigl(\state, \sched(\state, n), \state'\bigr)$ for $\state, \state' \in \states$ and $n \in \nats$, and $\rmat'(\cdot,\cdot) \defeq 0$ otherwise.
  \end{itemize}
  Let $\stopro^{\cmodel_\sched,\state_0}\colon (\samspace{\cmodel_\sched} \times \reals_{\geq 0}) \to (\states \times \nats)$ 
  be the stochastic process of the CTMC $\cmodel_\sched$, and let $f\colon (\states \times \nats) \to \states$ with $f(\state, n) \defeq \state$.
  The \emph{induced stochastic process} $\stopro^{\cmodel,\sched,\state_0}\colon (\samspace{\cmodel_\sched} \times \reals_{\geq 0}) \to \states$ of $\cmodel$ and $\sched$ starting in $\state_0$ is then defined as $\stopro^{\cmodel,\sched,\state_0}_t \defeq f \circ \stopro^{\cmodel_\sched,(\state_0,0)}_t$ for $t \in \reals_{\geq 0}$.
  Definitions for DTMDPs are likewise using $\pmat$ instead of $\rmat$.
\end{definition}

We equip our models with \emph{reward structures}, assigning values to states.
\begin{definition}
\label{def:reward-structure}
  A \emph{reward structure} for a stochastic process $\stopro\colon (\Omega \times \reals_{\geq 0}) \to \states$ or a CTMC or CTMDP with state set $\states$ is a tuple $(\crew, \frew)$ with $\crew\colon \states \to \reals_{\geq 0}$ and $\frew\colon \states \to \reals_{\geq 0}$.
  We call $\crew$ the \emph{cumulative reward rate}, and $\frew$ the \emph{final reward value}.
  We let $\frew^{\max} \defeq \max_{\state \in \states} \frew(\state)$, and $\crew^{\max} \defeq \max_{\state \in \states} \crew(\state)$.
\end{definition}

For CTMCs, the cumulative reward rate $\crew(\state)$ is the reward obtained per time unit for staying in state $\state$, until $\state$ is left or a given time bound $\timeb$ is reached.
The final reward value $\frew(\state)$ specifies the reward one obtains for being in state $\state$ at this time bound $\timeb$.
We are interested in the expected values of these numbers, as formalised in \refdef{value}.
For CTMDPs, we strive for the maximal (and analogously the minimal) value under all possible schedulers in the class we considered.
\begin{definition}
  \label{def:value}
  Given a time bound $\timeb \in \reals_{\geq 0}$, the \emph{value} of a stochastic process $\stopro\colon (\Omega \times \reals_{\geq 0}) \to \states$ with a reward structure $\rew = (\crew, \frew)$ is defined as $\mvalue(\stopro,\rew,\timeb) \defeq \expect \bigl[\int_0^\timeb \! \crew(\stopro_u) \, \diff u + \frew(\stopro_\timeb)\bigr]$.
  For a CTMC $\cmodel = (\states, \rmat)$ and $\state_0 \in \states$, we let $\mvalue(\cmodel, \state_0, \rew, \timeb) \defeq \mvalue(\stopro^{\cmodel,\state_0},\rew,\timeb)$.
  For a CTMDP $\cmodel = (\states, \acts, \rmat)$, the \emph{maximal value} (\emph{minimal value}) for $\state_0 \in \states$ is defined as $\mvalue^{\max}(\cmodel, \state_0, \rew, \timeb) \defeq \max_{\sched \in \schedshr} \mvalue(\stopro^{\cmodel,\sched,\state_0},\rew, \timeb)$ ($\mvalue^{\min}(\cmodel, \state_0, \rew, \timeb) \defeq \min_{\sched \in \schedshr} \mvalue(\stopro^{\cmodel,\sched,\state_0},\rew, \timeb)$).
\end{definition}
The interpretation of rewards and values depends on the model under consideration.
For instance, in a CTMC representing a chemical reaction, we might assign a final reward value of $n$ to state $\state$ if $\state$ contains $n$ molecules of a given species.
This way, the value of the CTMC represents the expected number of this species at a given point of time.

We do not explicitly consider impulse (instantaneous) rewards $\irew\colon (\states \times \states) \to \reals_{\geq 0}$ for CTMCs here, that is rewards obtained for moving from one state to another.
However, given cumulative reward rates $\crew$ and impulse rewards $\irew$, we can define cumulative reward rates $\crew'$ as $\crew'(\state) \defeq \crew(\state) + \sum_{\state' \in \states} \rmat(\state, \state') \cdot \irew(\state, \state')$.
For the properties under consideration, this new reward structure is equivalent to the one which uses impulse rewards (follows from~\cite[(6)]{KwiatkowskaNP06}).
\refdef{value} resembles the approach considered in a recent paper~\cite{BuchholzHHZ11}, where we maximised (minimised) the value over a more general class of schedulers than the one of \refdef{sched-hrs}.

An important specific value of a stochastic process is
the \emph{time-bounded reachability probability}. The computation
of this value is, for instance, necessary to model check the
time-bounded until property of the probabilistic logic CSL
(continuous stochastic logic)~\cite{BaierHHK03}.

Given a set of \emph{target states} $\target$, we can express the probability
to be in $\target$ \emph{at} time $t$ by using a reward structure with $\crew(\state) = 0$
for all $\state \in \states$, and $\frew(\state) = 1$ if $\state \in \target$
and $\frew(\state) = 0$ otherwise. For the probability to reach $\target$ \emph{within}
time $t$, we additionally modify the rate matrix $\rmat$ such that $\rmat(\state, \state) = \urate$,
and $\rmat(\state, \state') = 0$ for $\state' \neq \state$ if $\state \in \target$.
For CTMCs, the case to reach $\target$ within an interval $[a,b]$ with $0 < a < b < \infty$ can
be handled by two successive analyses~\cite[Theorem 3]{BaierHHK03}. Unbounded intervals
$[0,\infty)$ and $[a,\infty)$ can be handled similarly~\cite[Section 4.4]{KwiatkowskaNP07}.

The fact that \refdef{value} involves both final and cumulative rewards for CTMCs allows us to express the values at different points of time in the following way.
Assume we want to consider the cumulative reward rates $v_1, v_2, \ldots$ at consecutive time points $\timeb_1 = \delta_1, \timeb_2 = \timeb_1 + \delta_2, \ldots$
A short calculation then shows that $v_1 = \mvalue\bigl(\cmodel, \state, (\crew, 0), \delta_1\bigr)$, $v_2 = \mvalue\bigl(\cmodel, \state, (\crew, v_1), \delta_2\bigr)$, \dots{}
The formulation for the final reward at different points of time is likewise.

\begin{example}
  \label{exa:values}
  Reconsider the CTMC from \refexa{ctmc}, the CTMDP from \refexa{ctmdp}, and the ECTMC from \refexa{ectmc} all sketched in \reffig{examples}(a) through (c).
  Assume that for the CTMC we use a reward structure $\rew \defeq (\crew, \frew)$ with
  $\crew(\state_0) \defeq 0.0$, $\crew(\state_1) \defeq 0.25$, $\crew(\state_2) \defeq 0.5$, $\crew(\state_3) \defeq 0.75$, $\crew(\state_4) \defeq 1$,
  $\crew(\state_5) \defeq 1$, and $\frew(\cdot) \defeq 0$.
  Then we have that the reward value for $\state_0$ for a time bound of $\timeb \defeq 5$ is $\mvalue(\cmodel, \state_0, \rew, \timeb) \approx 2.70116$.

  Assume that, for the CTMDP and the ECTMC, we have $\crew(\astate_0) \defeq 0$, $\crew(\astate_1) \defeq 1$, $\crew(\astate_2) \defeq 1$, and $\frew(\cdot) \defeq 0$;
  and that we have $\timeb \defeq 5$. Then we have $\mvalue^{\max}(\cmodel, \astate_0, \rew, \timeb) \approx 4.30277$ for both models.
  Now assume that the reward values are $\crew(\astate_0) \defeq 0$, $\crew(\astate_1) \defeq 0.25$, $\crew(\astate_2) \defeq 1$, and $\frew(\cdot) \defeq 0$ instead.
  Then we have $\mvalue^{\min}(\cmodel, \astate_0, \rew, \timeb) \approx 1.82699$.
\end{example}


\subsection{PRISM's guarded command language}
\label{subsec:prism_gcl}

\noindent\PRISM~\cite{KNP11} is a widely used tool, which features a guarded command language to model CTMCs (among other classes).
For our purposes, it suffices to take a rather abstract view on the high-level modelling language used by this tool.
\begin{definition}
\label{def:prism-model}
  A \emph{\PRISM model (PM)} is a tuple $\prismModel = (\prismVars, \prismInit, \prismCmds, \prismSucc, \prismCRew, \prismFRew)$.
  Here, $\prismVars$ is a set of Boolean \emph{variables}. With $\prismInit\colon \prismVars \to \{0,1\}$,
  we denote the \emph{initial state}, and $\prismCmds$ is a finite set of \emph{commands}.
  Let $\states_{\prismVars} \defeq \bigl\{s\colon\prismVars \to \{0, 1\}\bigr\}$ be the set of variable assignments for $\prismVars$.
  The \emph{cumulative reward rate} is a function $\prismCRew\colon \states_{\prismVars} \to \reals_{\geq 0}$ as is the \emph{final reward value} $\prismFRew\colon \states_{\prismVars} \to \reals_{\geq 0}$.
  The \emph{successor function} is a partial function of the form $\prismSucc\colon (\states_{\prismVars} \times \prismCmds) \partialto (\states_{\prismVars} \times \reals_{> 0})$.
  We define $\prismNSucc(\state, \prismCmd) \defeq \state'$ if $\prismSucc(\state, \prismCmd) = (\state', \lambda)$ for some $\lambda \in \reals_{> 0}$.
  We also let
\begin{equation*}
  \prismSucc(\state) \defeq \Bigl\{(\state', \lambda) \,\Big|\,\exists \prismCmd . \prismNSucc(\state, \prismCmd) = \state' 
  \wedge \lambda = \hspace*{-2em} \sum_{\lambda'; \exists \prismCmd' . (\state', \lambda') = \prismSucc(\state, \prismCmd')}\hspace*{-2em}\lambda'\hspace*{1.5em} \Bigr\}.
\end{equation*}
  Further, $\prismNSucc(\state) \defeq \bigl\{\state' \,\big|\, \exists \lambda . (\state', \lambda) \in \prismSucc(\state) \bigr\}$.
  Let $\prismNSucc^0 \defeq \prismInit$, and $\prismNSucc^{i+1} \defeq \{\prismNSucc(\state) \mid \state \in \prismNSucc^{i}\}$.
  The set of \emph{reachable states (state space)} is $\states_{\prismModel} \defeq \bigcup_{i=0}^\infty \prismNSucc^{i}$.
  We require that there is $\urate(\prismModel) > 0$ such that $\urate(\prismModel) = \sum \bigl\{ \lambda \,\big|\, \exists \state' . (\state',\lambda) \in \prismSucc(\state) \bigr\}$ for all $\state \in \states_\prismModel$ (where the latter is a multiset).
\end{definition}
The complete \PRISM syntax also defines models consisting of several \emph{modules}, that is, sets of guarded commands, which may synchronise or interleave.
However, as the semantics of a model with several modules is defined as one with a single module, a single set of commands suffices.
\PRISM also allows us to specify commands with several pairs of successors $(\state_1',\lambda_1), \ldots, (\state_n', \lambda_n)$.
For PMs describing CTMCs, such a command is equivalent to a set of $n$ commands $\prismCmd_i$ in the above form.
Each of them must be activated (defined) in the same states as the original command, and we then have $\prismSucc(\prismCmd_i) = (\state_i', \lambda_i)$.
The bounded integers \PRISM supports can be represented by a binary encoding.
Impulse rewards can be transformed to cumulative rewards, as discussed for CTMCs.
For models in which there is no $\urate(\prismModel)$ with the required property, we can add an extra command to increase the self-loop rate where necessary.

The formal semantics of a PM is as follows.
\begin{definition}
\label{def:induced-ctmc-rewards}
  Consider a PM $\prismModel = (\prismVars, \prismInit, \prismCmds, \prismSucc,$ $\prismCRew, \prismFRew)$.
  The \emph{induced CTMC} is $\cmodel_\prismModel \defeq (\states_\prismModel, \rmat)$ 
such that for all $\state, \state' \in \states_\prismModel$ we have
$\rmat(\state, \state') \defeq \lambda$ if $(\state',\lambda) \in \prismSucc(\state)$, and $\rmat(\state, \state') \defeq 0$ if no such tuple exists.
  The \emph{induced reward structure} is $\rew_\prismModel \defeq (\prismCRew, \prismFRew)$.
\end{definition}

In \refsec{algorithms}, we will abstract CTMCs into ECTMCs.
To do so, we will subsume several concrete states of a CTMC to abstract states of an ECTMC.
\begin{definition}
  \label{def:partitioning}
  Given a PM $\prismModel$, a \emph{partitioning} of the state space
  $\states_{\prismModel}$ is a finite ordered set $\apart = \langle\astate_0,\ldots,\astate_{n-1}\rangle$ of
  non-empty, pairwise disjoint subsets of $\states_{\prismModel}$ such that
  $\states_{\prismModel} = \bigcup_{i=0}^{n-1}\astate_i$.  
\end{definition}

\begin{example}
  \label{exa:prism}
  In \reffig{examples}(d) we give an example of a \PRISM model.
  The description is given in the textual form which is used by the tool itself.
  The first line states that the model is a CTMC.
  Then, a single module with the name \texttt{example} is declared.
  In this module, there are two variables \texttt{n}, and \texttt{m} with a specified variable range, and initial value.
  Afterwards, the successor function is given in terms of guarded commands of the form \texttt{<guard> -> <rate> : <successor>}.

  The induced CTMC is the model in \reffig{examples}(a) (for readability, we have left out the commands that lead to the self-loops in this model).
  Here, $\state_0$ corresponds to $\mbox{\texttt{n}}=0,\,\mbox{\texttt{m}}=1$, $\state_5$ corresponds to 
  $\mbox{\texttt{n}}=2,\,\mbox{\texttt{m}}=1$, and the other $\state_i$ correspond to $\mbox{\texttt{n}}=1,\,\mbox{\texttt{m}}=i$.

  The reward structure is given below the module definition.
  \PRISM only supports either final or cumulative rewards, but not both at the same time.
  Thus, either the final or the cumulative reward part is zero.
  Whether the reward specified this way shall denote the final 
  or cumulative reward is decided by using a formula specification.

  As an example partitioning, we can consider $\apart=\langle\astate_0,\astate_1,\astate_2\rangle$ with
  $\astate_0 = \{\state_0\}$, $\astate_1 = \{\state_1, \state_2, \state_3, \state_4\}$, and $\astate_2 = \{\state_5\}$.
\end{example}

\subsection{Binary decision diagrams}
\label{subsec:obdds}

\noindent\emph{Binary decision diagrams}~\cite{Bry86} are an efficient tool to
symbolically represent structures which are too large to be represented in an
explicit form.
\begin{definition}
  \label{def:bdd}
  We fix a finite ordered set $\bddVars \defeq \langle \bddVar_1, \ldots,
  \bddVar_m\rangle$ of \emph{Boolean variables}.
  A \emph{binary decision diagram} (BDD) is a rooted acyclic directed graph
  $\bdd$ with node set $\bddNodes$, and root node $\bddNodeInit$. There are two
  types of nodes in $\bddNodes$: terminal nodes, and non-terminal nodes.
  \emph{Terminal} nodes $\bddNode$ do not have out-going edges, and are
  labelled with a value $\bddNV(bddNode)\in\{0,1\}$. The remaining nodes are
  \emph{non-terminal} nodes $\bddNode\in\bddNodes$, which have exactly two successor nodes, denoted by
  $\bddHi(\bddNode)$ (\emph{high} successor), and $\bddLo(\bddNode)$ (\emph{low} successor).
  Non-terminal nodes $\bddNode$ are labelled with a variable
  $\bddNV(\bddNode)\in\bddVars$.
\end{definition}
A \emph{variable valuation} is a function $v\colon \bddVars \to \{0,1\}$.
We denote the set of all variable valuations by $\bddEval$.
Each valuation $v$ induces a unique path in the BDD from the root node to a
terminal node.
At a non-terminal node $\bddNode$, we follow the edge to $\bddHi(\bddNode)$ if
$v\bigl(\bddNV(\bddNode)\bigr) = 1$, and the edge to $\bddLo(n)$ if
$v\bigl(\bddNV(\bddNode)\bigr)=0$. The function $\sem{\bdd}:\bddEval\to\{0,1\}$
represented by a BDD $\bdd$ returns for a variable valuation $v$ the value of
the terminal node reached by following the path induced by $v$.
\begin{definition}
  \label{def:reduced_ordered}
  A BDD is \emph{ordered} if for all non-terminal nodes $n$ the following condition
  holds.
  Either $\bddHi(\bddNode)$ is a terminal node, or $\bddNV(\bddNode) <
  \bddNV\bigl(\bddHi(\bddNode)\bigr)$, and the same for $\bddLo(\bddNode)$. A BDD is
  \emph{reduced} if all sub-BDDs rooted at the different nodes of the
  BDD represent distinct functions. Reduced and ordered BDDs are called
  \emph{OBDDs}.
\end{definition}

The OBDD for the constant 0 (or 1) function, which consists of a single 
terminal node labelled with a 0 (or a 1), is denoted in the sequel 
by $\bddZero$ (or $\bddOne$, respectively).

OBDDs are a canonical representation (up to isomorphism) of
arbitrary functions $f\colon\bddEval\to\{0,1\}$~\cite{Bry86}. In the following, we
will only use OBDDs. For more details on (O)BDDs, we refer the reader
to~\cite{Bry86,Weg00}.

OBDDs support a wide number of operations like the Boolean operations $\wedge$,
$\vee$, and $\neg$. Given two ordered sets $\bddVars_1 = \langle \bddVar_{i_1}, \ldots, \bddVar_{i_n} \rangle$ and
$\bddVars_2 = \langle \bddVar_{j_1}, \ldots, \bddVar_{j_n} \rangle$ of
Boolean variables, by $\bdd' = \bdd[\bddVars_1 / \bddVars_2]$ we denote the OBDD
which results from renaming the variables in $\bddVars_1$ to the corresponding
variables in $\bddVar_2$.
For $\bddVars' \subseteq \bddVars$ and OBDD $\bdd$, we let 
$\sem{\exists\bddVars'.\bdd}(v) = \bigvee \bigl\{ \sem{\bdd}(v') \,\big|\, \forall \bddVar
\notin \bddVars' . v'(x) = v(x) \bigr\}$ be the existential quantification of the
variables in $\bddVars'$.

We can use OBDDs to represent PMs in a symbolic form, if we leave out the stochastic aspects.
\begin{definition}
  \label{def:bdd-repr}
  Consider a PM $\prismModel = (\prismVars, \prismInit, \prismCmds, \prismSucc,$ $\prismCRew, \prismFRew)$.
  The \emph{OBDD representation} of $\prismModel$ is a tuple $\bdd_\prismModel \defeq \bigl(\prismVars, \prismVars', \bddInit, \{ \bddSucc_\prismCmd \}_{\prismCmd \in \prismCmds}\bigr)$.
  There, $\prismVars$ and $\prismVars'$ are sets of Boolean variables with $\prismVars \cap \prismVars' = \emptyset$ such that there is a one-to-one mapping between variables $\prismVar \in \prismVars$ and $\prismVar' \in \prismVars'$.
  Further, $\bddInit$, and $\bddSucc_\prismCmd$ are OBDDs over the variables $\prismVars$, and $\prismVars \cup \prismVars'$, respectively.
  We require that $\sem{\bddInit}(v) = 1$ iff $\prismInit(\prismVar) = v(\prismVar)$ holds for all $\prismVar \in \prismVars$.
  For all $\bddSucc_\prismCmd$, we require $\sem{\bddSucc_\prismCmd}(v) = 1$ iff for all $\prismVar \in \prismVars$ it is true
  that $v(\prismVar) = \state(\prismVar)$, $v(\prismVar') = \state'(\prismVar)$, and $\prismNSucc(\state, \prismCmd) = \state'$.
  By $\bddSucc$, we denote the OBDD such that $\sem{\bddSucc} = \bigvee_{\prismCmd \in \prismCmds} \sem{\bddSucc_\prismCmd}$.
\end{definition}

OBDDs can also be used to symbolically represent a partitioning of the state space of a PM.
Let $\apart = \langle\astate_0,\ldots,\astate_{n-1}\rangle$ be a partitioning of the PM $\prismModel = (\prismVars, \prismInit, \prismCmds, \prismSucc, \prismCRew, \prismFRew)$.
The idea is to assign to each block $\astate_i$ of $\apart$ a unique block number $i$, and to use a binary representation of $i$, which is encoded using $k = \lceil \log_2 n\rceil$ novel BDD variables $\avars = \langle\avar_0,\ldots,\avar_{k-1} \rangle$.
\begin{definition}
  \label{def:bdd-partitioning}
  The \emph{OBDD representation} of $\apart = \langle\astate_0,\ldots,\astate_{n-1}\rangle$ is the OBDD $\bdd_\apart$ over the variables $\prismVars\uplus\avars$, where $\avars = \langle\avar_0,\ldots,\avar_{k-1} \rangle$ with $k = \lceil \log_2 n\rceil$.
  We require that $\sem{\bdd_\apart}(v) = 1$ iff there is $\state \in \states_\prismModel$ such that, for all $\prismVar \in \prismVars$,
  we have $v(\prismVar) = \state(\prismVar)$; and there is $\astate \in \apart$ such that $\state \in \astate$, and for all $\avar \in \avars$ we have $v(\avar) = \astate(\avar)$.
  With $\astate_i(\avar_j) \defeq (i\ \mathrm{div}\ 2^j) \mod 2$, we denote the value of variable $\avar_j \in \avars$ in the binary encoding of the block number $i$ of $\astate_i \in \apart$.
  With $\bdd_\astate$, we denote the OBDD such that $\sem{\bdd_\astate}(v) = 1$ iff $v$ represents a state of $\astate$.
\end{definition}

There are several alternative OBDD-based partition representations available.
One possibility is to use an MTBDD with the block numbers in the leaves.
However, the algorithms of the tool~\SIGREF~\cite{wimmer-et-al-atva-2006}, which we use
for refining partitions, require us to represent sets of block numbers within the BDD, 
which is easily possible using our encoding. To do so, the authors of \cite{Santos13}
encode such sets as products of prime numbers. However, in our application this would yield numbers which are too 
large to fit into the standard 32- or 64-bit data types. Therefore this approach is not directly 
possible, and requires technical tricks which might affect the efficiency of the method.
Another possibility is to use one OBDD for each abstract state. In this representation,
finding the abstract state containing a given concrete state is more expensive
(in $O(|\prismVars|\cdot |\apart|)$ instead of $O(|\prismVars|+|\apart|)$).
The logarithmic partition encoding proposed by Derisavi~\cite{Der07a} has certain advantages 
regarding memory consumption, but partition refinement is much more expensive regarding 
computation time than the representation we use~\cite{wimmer-et-al-perfeval-2010}.
Finally, Bouali and de Simone~\cite{BdS92} represent the corresponding equivalence
relation for bisimulation computation. In previous work,~\cite{wimmer-et-al-atva-2006},
we have observed that this representation is often larger than the one used by \SIGREF, 
slowing down the computations.

Regarding the variable order of $\prismVars\uplus\avars$, we assume in
the following that all variables in $\prismVars$ precede all variables
in $\avars$. This ordering leads to more efficient algorithms for accessing the
block number of a given state.

\begin{example}
  Reconsider the \PRISM model from \refexa{prism}.
  We can encode each of the variables \texttt{n} and \texttt{m} by two binary variables $\mbox{\texttt{n}}_0$, $\mbox{\texttt{n}}_1$ and $\mbox{\texttt{m}}_0$, $\mbox{\texttt{m}}_1$ by using the binary representations
  $\mbox{\texttt{n}}=0 \leftrightarrow (\mbox{\texttt{n}}_1,\mbox{\texttt{n}}_0)=(0,0)$,
  $\mbox{\texttt{n}}=1 \leftrightarrow (\mbox{\texttt{n}}_1,\mbox{\texttt{n}}_0)=(0,1)$,
  $\mbox{\texttt{n}}=2 \leftrightarrow (\mbox{\texttt{n}}_1,\mbox{\texttt{n}}_0)=(1,0)$,
  $\mbox{\texttt{m}}=1 \leftrightarrow (\mbox{\texttt{m}}_1,\mbox{\texttt{m}}_0)=(0,0)$,
  $\mbox{\texttt{m}}=2 \leftrightarrow (\mbox{\texttt{m}}_1,\mbox{\texttt{m}}_0)=(0,1)$,
  $\mbox{\texttt{m}}=3 \leftrightarrow (\mbox{\texttt{m}}_1,\mbox{\texttt{m}}_0)=(1,0)$,
  $\mbox{\texttt{m}}=4 \leftrightarrow (\mbox{\texttt{m}}_1,\mbox{\texttt{m}}_0)=(1,1)$.
  To encode the partitioning of the previous example ($\astate_0 = \{\state_0\}$, $\astate_1 = \{\state_1, \state_2, \state_3, \state_4\}$, 
  and $\astate_2 = \{\state_5\}$), we enumerate $\astate_i$ by introducing new binary variables $k_0$, $k_1$.
  We sketch the encoding in \reffig{examples}(g).
  The OBDD terminal nodes are given as squares at the bottom of the diagram.
  Non-terminal nodes are given as circles, labelled with their variables.
  The $\bddHi$ successors are connected by solid lines, whereas $\bddLo$ successors are connected by dashed ones.
  For readability, we leave out the connections to the $0$ terminal node.
\end{example}

\section{Algorithms}
\label{sec:algorithms}
\noindent In this section, we first describe an algorithm to
approximate minimal and maximal values of CTMDPs. Afterwards, we
describe how to obtain an ECTMC from a PM, such that its induced CTMDP
is a valid abstraction (cf. \refpro{corr-abs}) of the CTMC semantics
of the PM. We provide an algorithm which computes an ECTMC
over-approximation of a PM given in an OBDD representation. Using the
first algorithm, we can obtain intervals from this abstraction which
are guaranteed to bound the actual value (cf.~Definition~\ref{def:value})
of the CTMC from above and below.

\subsection{Computing Reward Values for CTMDPs}

\noindent Let $\phi_\lambda(i) \defeq \lambda^ie^{-\lambda}/(i!)$ denote 
the probabilities of a Poisson distribution with parameter $\lambda$, and let 
$\psi_\lambda(i) \defeq \sum_{j=i+1}^\infty \phi_\lambda(j) = 1 - \sum_{j=0}^i\phi_{\lambda}(j)$.

The algorithm to compute the maximal values of CTMDPs is given in \refalg{opt-reach}.
The input is a CTMDP $\cmodel$ with reward structure $\rew = (\crew, \frew)$,
and the precision $\precision>0$ up to which the values are to be computed.
The algorithm for the minimum is likewise, replacing $\max$ by $\min$
in Line~\ref{lin:main-maximise}.
\begin{algorithm}
  \caption{\label{alg:opt-reach}Compute maximal values for $\cmodel = (\states, \acts, \rmat)$, $\rew = (\crew, \frew)$ up to $\precision$.}
  \DontPrintSemicolon
  let $k$ s.\,t. $\sum_{i=0}^k\psi_{\urate \timeb}(i) > \urate \timeb -
  \precision \urate / (2\crew^{\max})
  \wedge \psi_{\urate \timeb}(k) \cdot \frew^{\max} <
  \precision/2$ \label{lin:choice_k} \;
  $\cmodel' = (\states, \acts, \pmat) \gets \emb(\cmodel)$ \;
  \lForAll{$\state \in \states$} {\label{lin:init}%
    $q_{k+1}(\state) \gets 0$ \;
  }
  \ForAll{$i=k,k-1,\ldots,0$} { \label{lin:main-start}
    \ForAll{$\state \in \states$} {
      $m \gets \max\limits_{\act\in\acts(s)} \sum\limits_{\state' \in S} \pmat(s,\act,s') q_{i+1}(s')$ \label{lin:asg-max} \; \label{lin:main-maximise}
      $q_i(\state) \gets m + \phi_{\urate \timeb}(i) \cdot \frew(\state) +
      \psi_{\urate \timeb}(i) \cdot \crew(\state)/\urate$ \;
      \label{lin:last-inner}
    }
  }
  \Return $q_0$
\end{algorithm}

The choice of $k$ in line~\ref{lin:choice_k} of \refalg{opt-reach} 
is based on the following lemma, which is proven in the Appendix.
\begin{lemma}
  \label{lem:precbound}
  Given a CTMDP $\cmodel = (\states, \acts, \pmat)$ with a reward structure $\rew = (\crew, \frew)$, a precision $\precision > 0$, and $k$ such that
  \[
  \sum_{n=0}^k\psi_{\urate \timeb}(n) > \urate \timeb - \frac{\precision \urate}{2\crew^{\max}} \mbox{ and } \psi_{\urate \timeb}(k) \cdot \frew^{\max} < \frac{\precision}{2},
  \]
  then for all schedulers $\sched \in \schedscr$, and all $\state_0 \in \states$, we have
\begin{equation*}
\begin{split}
  & \sum_{i=k+1}^\infty \left(\phi_{\urate \timeb}(i) \sum_{\state \in \states} \tprob^{\emb(\cmodel),\sched}(\state_0, i, \state) \frew(\state)\right. \\
  & + \left.\psi_{\urate \timeb}(i) \sum_{\state \in \states} \tprob^{\emb(\cmodel),\sched}(\state_0, i, \state)
    \frac{\crew(\state)}{\urate}\right) < \precision .
\end{split}
\end{equation*}
Here, $\tprob^{\emb(\cmodel),\sched}(\state_0, i, \state)$ is the probability of being in state $\state$ of $\emb(\cmodel)$ in step $i$ if having started in $\state_0$ when using scheduler $\sched$.
By $\schedscr$, we denote the set of schedulers which extend $\schedscd$ by randomised choice over the actions.
\end{lemma}
The requirement on the actions in \refdef{ctmdp} assures that the
maximum in \refalg{opt-reach} (line~\ref{lin:asg-max}) exists. We can also directly apply this
algorithm on ECTMCs without constructing the uncountably large induced
CTMDPs. The crucial part here is the optimisation over the
uncountable actions, which can be done using a slight adaptation of
methods from~\cite[Chapter 4.1]{Klink10}. There, optimising the
assignment of successor rates with restrictions given by lower and
upper bounds is already described. 
For each $\state \in \states$, and for each $\hat{\act}$ such that $\ivall(\state,\hat{\act})$ (and thus $\ivalu$) is defined, we can apply the method described in \cite[Chapter 4.1]{Klink10}, thus to find the optimal $v_{\hat{\act}}\colon \states \to [0,\urate]$ for this $\hat{\act}$.
Afterwards, we choose the optimal $v_{\hat{\act}}\colon \states \to [0,\urate]$ among all $\hat{\act}$, which is easy as there are only a finite number.
\begin{proposition}
  \label{pro:algo-correctness}
  Let $\cmodel = (\states, \acts, \rmat)$ be a CTMDP with reward structure $\rew = (\crew, \frew)$.
  Then, there exists $\sched \in \schedscd$ such that $\mvalue^{\max}(\cmodel, \state_0, \rew, \timeb) = \mvalue(\stopro^{\cmodel,\sched,\state_0},\rew, \timeb)$ for all $\state_0 \in \states$.
  Further, the return value $q_0$ of \refalg{opt-reach} is such 
  that $|\mvalue^{\max}(\cmodel, \state_0, \rew, \timeb) - q_0(\state_0)| < \precision$.
\end{proposition}
\begin{proofsketch}
  At first, we show that we can simulate each history-dependent
  randomised scheduler by a randomised counting scheduler (CR).
  In contrast to CD, these schedulers may be randomised.
  However, as for CDs, decisions of these schedules only depend on the number of steps which have passed rather than on the full history.
  The fact that CR can simulate HR schedulers allows us to use a result about discrete-time Markov chains.
  Next, we show that \refalg{opt-reach} cannot yield values which are
  larger than the maximal value resulting from such a CR. Then, we show
  that the algorithm does not return values which are larger than the
  value obtained by any CR plus the specified precision.

  Looking at the decisions the algorithm takes at \reflin{asg-max}, we
  can reconstruct a prefix of the decisions of a CD. By letting the
  precision approach $0$, we can show that there is indeed a complete CD
  yielding the same value.
\end{proofsketch}
  The full proof can be found in the Appendix.

\refalg{opt-reach} generalises an approach from a previous paper about
time-bounded reachability~\cite{BaierHKH05} using results by
Kwiatkowska et al.~\cite{KwiatkowskaNP06}. Its correctness also
proves that deterministic counting schedulers suffice to obtain
optimal values, because the algorithm implicitly computes such a
scheduler.

It is also related to an earlier work in queueing theory
\cite{Lippman76}, which is however different in a number of ways. The
target there was to obtain approximations for a more general class of
schedulers of CTMDPs than we need here, and thus does not consider
maxima over HRs explicitly. It assumes a fixed maximal number of steps
to happen in the uniformised DTMDP, rather than deriving the necessary
number, as we do in our algorithm. \cite{Lippman76} is also more
involved with models featuring a particular structure rather than
computing conservative bounds on properties of CTMCs.

The next proposition states how CTMDPs can be used to over-approximate
CTMCs.
\begin{proposition}
  \label{pro:corr-abs}
  Let $\cmodel = (\states, \rmat)$ be a CTMC with reward structure
  $(\crew, \frew)$, and let $\apart = \langle\astate_0,\ldots,\astate_{n-1}\rangle$ be a partitioning
  of $\states$. Consider the CTMDP $\cmodel' \defeq (\apart, \acts, \rmat')$
  where for each $\astate \in \apart$ and $\state \in \astate$
  we find $\act_\state \in \acts$ such that for all $\astate' \in
  \apart$ we have
  $\rmat'(\astate,\act_\state,\astate') \defeq \sum_{\state' \in \astate'} \rmat(\state, \state')$.
  Further, consider a reward
  structure $(\crew', \frew')$ such that for all $\astate \in \apart$
  it is true that $\crew'(\astate) \geq \max_{\state \in \astate}
  \crew(\state)$, and $\frew'(\astate) \geq \max_{\state \in
    \astate} \frew(\state)$. Then, for all $\astate_0 \in \apart$
  and $\state_0 \in \astate_0$, we have $\mvalue(\cmodel, \state_0,
  \rew, \timeb) \leq \mvalue^{\max}(\cmodel', \astate_0, \rew',
  \timeb)$.
\end{proposition}
\begin{proofsketch}
  We can construct a scheduler $\sched$ such that the embedded DTMDP
  of $\cmodel'$ mimics the behaviour of the embedded DTMC of
  $\cmodel$, so that in each step the probability to be in a given
  abstract state $\astate$ is the sum of the probabilities of being in
  a state $\state$ of $\cmodel$ with $\state \in \astate$.
  By the definition of reward structures, the value obtained in
  $\cmodel'$ using $\sched$ is as least as high as the value in
  $\cmodel$.
  As the maximal value in $\cmodel'$ is at least as high as the one
  using $\sched$, the result follows.
\end{proofsketch}
  The full proof can be found in the Appendix.

We remark that Algorithm~\ref{alg:opt-reach} can only compute maximal bounds up to $\precision$.
Thus, when applying it to compute upper bounds on the values of CTMCs, one has to add $\precision$ 
to get a number which is guaranteed to bound the value of the original model from above.
Alternatively, one could apply bounding semantics~\cite{ZhaoC13}.

As mentioned before, CSL bounded-until properties can be expressed
using rewards. Their probabilities can thus also be bounded using
\refalg{opt-reach}.

We remark that, for the case of intervals $[a,b]$ or $[a,\infty)$,
the successive application of the algorithm only bounds the value
in the CTMC that has been abstracted. Assume that one divides the
interval into several parts, and successively applies the algorithm
for each of these parts by using the result of the previous application
as the instantaneous reward of the next application. Doing so allows
us to obtain values at different points of time. However, the last minimal
(or maximal, respectively) value is not guaranteed to be identical to
the value which one would have obtained by a single analysis of the
whole interval.

\begin{figure}
\centering
\includegraphics{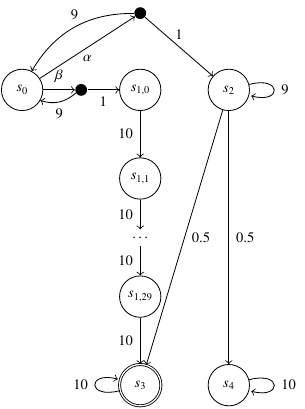}
\caption{\label{fig:nountil}Example demonstrating that successive applications of
  Algorithm~\ref{alg:opt-reach} to sub-intervals are not equivalent to a single
  application on the whole interval.}
\end{figure}
\begin{example}
Consider the CTMDP $\cmodel$ in \reffig{nountil}, which is adapted
from previous publications~\cite{ZhangN10,BuchholzHHZ11}.
The only difference is that non-uniform CTMDPs were used previously,
i.\,e., CTMDPs in which the sum of leaving rates may be different for each state.
Here, we have increased the self-loop rates thus to obtain a uniformisation rate of $\urate = 10$.
States of the model are given as circles, in which the state name is written.
If there is more than one activated action in a state, we draw all of them as small black circles connected to the corresponding states.
Non-zero transition rates are then drawn starting in the corresponding black circle.
In case there is just one possible action in a state, we directly draw the transitions from this state.
Rates with value $0$ are left out.

We assign a cumulative reward rate of $0$ to each state.
The final reward is $0$ for all states except for $\state_3$, where it is $1$.
Intuitively, if we want to maximise the value, the optimal choice of the action depends on the amount of time left.
If much time is left, it is best to choose $\beta$ in $\state_0$, because this action leads to a sequence of states which, given an infinite amount of time, always reaches $\state_3$.
If little time is left, it is better to choose $\alpha$, because then $\state_3$ can be reached quickly, although there is a significant chance that this state will not be reached at all.

With this model and reward structure, we can compute the probabilities of the CSL formula
$\mathcal{P}\left( \mathit{true}\ \mathcal{U}^{[0,4]}\ \state_3 \right)$
by computing the value for $\timeb = 4$.
Because the state $\state_3$ is absorbing, this value is also the probability of the interval-bounded until property
$\mathcal{P}\left( \mathit{true}\ \mathcal{U}^{[1,4]}\ \state_3 \right)$.

In this special case, we can thus compute this probability using the method developed in this paper, or by the previous algorithm by Baier et al.~\cite{BaierHKH05}.
For $\state_0$, this value is $\mvalue^{\max}(\cmodel, \state_0, (0, \frew), 4) \approx 0.659593$.

Now we apply two consecutive analyses to bound the interval-bounded reachability probability.
For this procedure, we first compute $v(\cdot) \defeq \mvalue^{\max}(\cmodel, \state_0, (0,\frew), \cdot, 3)$,
and afterwards we consider $v'(\cdot) \defeq \mvalue^{\max}(\cmodel, \state_0, (0,v), \cdot, 1)$.
We now have $v'(\state_0) \approx 0.671162$, which shows that this value is an upper bound for the reachability probability in the original model.
It is indeed between the value obtained using HRs as discussed previously and the one obtained using time-dependent schedulers~\cite{ZhangN10,BuchholzHHZ11}. 
This value is larger than the one considered in the last paragraph, which shows that a consecutive analysis does not yield the maximum interval-bounded reachability probability over all HRs in a given CTMDP.
The reason that this outcome happens is that, by dividing the analysis into two parts, the schedulers of the two analyses can change their decisions more often,
and obtain more information than they are supposed to have.
Instead of only having the information that the objective is to
optimise the reward until $\timeb = 4$, it is now also known whether
$\timeb \leq 1$ or not.
\end{example}

From the discussion of the values of consecutive time points 
$\timeb_1 = \delta_1, \timeb_2 = \timeb_1 + \delta_2, \ldots$, 
it follows that we can also use the algorithm to compute bounds for
them in an efficient way. Instead of doing analyses with time bounds
$\timeb_1, \timeb_2, \ldots$, we only have to do analyses with
values $\delta_1, \delta_2, \ldots$, which might be much smaller than
$\timeb_1, \timeb_2, \ldots$ Thus, the fact that the algorithm
allows us to handle final and cumulative rewards at the same time has the
potential to speed up such a series of analyses.

\subsection{Abstracting \PRISM Models}

\noindent To take advantage of \refpro{corr-abs}, we want to avoid actually
constructing the CTMC to be abstracted. Doing so allows us to handle
models which are too large to be handled in an explicit-state form.
For this approach, we can use non-probabilistic model checkers which feature a
guarded-command language, like \NUSMV~\cite{CimattiCGGPRST02}. Such a
tool can work with an OBDD-based representation of PMs as in
\refdef{bdd-repr}, and compute the set of reachable states. We can
then specify some OBDDs representing \emph{predicates}, i.\,e., sets
of concrete states. These sets can be used to split the state space
by subsuming all concrete states that are contained in the same
subset of predicates, thus to obtain an OBDD partitioning as in
\refdef{bdd-partitioning}.

Next, we consider the ECTMC abstraction of a given \PRISM model.
\begin{definition}
  \label{def:cmc-abs}
  Consider a PM $\prismModel = (\prismVars, \prismInit, \prismCmds, \prismSucc,$ $\prismCRew, \prismFRew)$ 
  with induced CTMC $\cmodel = (\states, \rmat)$, and a partitioning 
  $\apart = \langle \astate_0, \ldots, \astate_{n-1}\rangle$ of its state space.
  The \emph{ECTMC abstraction} of $\prismModel$ is defined as
  $\cmodel \defeq (\apart, \widehat{\acts}, \ivall, \ivalu)$
  with $\widehat{\acts} \defeq \{ \hat{\act}\colon \prismCmds \partialto \apart \}$.
  We let $A(\astate, \hat{\act})$ denote the set of all $\state \in \astate$ such that 
  $\dom(\prismSucc(\state, \cdot)) = \dom(\hat{\act})$ (that is, the domains of 
  the two partial functions agree), and for all applicable $\prismCmd \in \prismCmds$ 
  we have $\prismNSucc(\state, \prismCmd) \in \hat{\act}(\prismCmd)$.
  We then choose the domain of $\ivall$ and $\ivalu$ such that, for all $\astate \in \apart$, we have
  \[
  \dom(\ivall(\astate, \cdot))
  \defeq \dom(\ivalu(\astate, \cdot))
  \defeq \{ \hat{\act} \in \widehat{\acts} \mid A(\astate, \hat{\act}) \neq \emptyset \} .
  \]
  Then, for $\astate,\astate' \in \apart$, and $\hat{\act} \in \dom(\ivalu(\astate, \cdot))$ we define
  \[
  \ivall(\astate, \hat{\act})(\astate') \defeq \min_{\state \in A(\astate, \hat{\act})} \sum_{\substack{(\state', \lambda) \in \prismSucc(\state), \\ \state' \in \astate'}} \lambda,
  \]
  and accordingly $\ivalu$ using $\max$.
  The abstract reward structure $\rew \defeq (\crew, \frew)$ is defined as $\crew(\astate) \defeq \max_{\state \in \astate} \prismCRew(\state)$,
  and $\frew(\astate) \defeq \max_{\state \in \astate} \prismFRew(\state)$.
\end{definition}
By construction, the CTMDP semantics of the ECTMC fulfils the
requirements of \refpro{corr-abs} for a correct abstraction of the
CTMC semantics of the \PRISM model. It is also monotone in the sense
that, by using a refined partitioning, we cannot obtain worse bounds
than with a coarser partitioning.
\begin{proposition}
  \label{pro:monotone}
  Consider a PM $\prismModel = (\prismVars, \prismInit, \prismCmds, \prismSucc,$ $\prismCRew, \prismFRew)$ 
  with a partitioning $\apart = \langle\astate_0, \ldots, \astate_{n-1}\rangle$ 
  of the state space of its induced CTMC, and a further partitioning 
  $\apart' = \langle \astate_0', \ldots, \astate_{m-1}'\rangle$ such that for each 
  $\astate_t \in \apart$ we find $\astate_{t,1}', \ldots, \astate_{t,u}' \in \apart'$ 
  such that $\astate_t = \bigcup_{j=1}^u \astate_{t,j}'$.

  Then, for two ECTMC abstractions 
  $\cmodel \defeq (\apart, \widehat{\acts}, \ivall, \ivalu)$ and 
  $\cmodel' \defeq (\apart', \widehat{\acts'}, \ivall', \ivalu')$ 
  with corresponding reward structures $\rew$ and $\rew'$,
  we have $\mvalue^{\max}(\cmodel, \astate_t, \rew, \timeb) \geq \mvalue^{\max}(\cmodel', \astate_{t,j}', \rew', \timeb)$.
\end{proposition}

\begin{proofsketch}
  We can show that, for an arbitrary $\precision > 0$, we have that $\mvalue^{\max}$ of a state $\astate$ 
  of the original partition plus $\precision$ is at least equal to the value of a state $\astate'$ of the 
  refined partition for which we have $\astate' \subseteq \astate$.
  This result implies that the same holds for $\precision = 0$, which means that the value of 
  a state of the refined partition cannot be higher than the one of the original abstraction.

  We use the fact that we can apply \refalg{opt-reach} to compute values up to any precision $\precision > 0$.
  Consider the runs of the algorithm on the original, and refined partitioning.
  Before the execution of the main loop at \reflin{main-start}, we have that 
  $q_{k+1}(\astate) = q_{k+1}(\astate') = 0$. For each execution of the main loop, 
  we have a maximising decision in $\astate'$, leading to $v_i(\astate')$ to be 
  added to $q_{i+1}(\astate')$ to obtain $q_i(\astate')$.
  We can construct a decision for $\astate$ such that $v_i(\astate')\leq v_i(\astate)$.
  This result means that, in each iteration of the main loop, the $q_i$ of $\astate'$ 
  can never become larger than the one of $\astate$. Thus, after the termination of 
  the main loop and the algorithm, the value obtained for the coarser abstraction is 
  at least as high as the one in the refined abstraction.
\end{proofsketch}
  The full proof can be found in the Appendix.

\begin{example}
  \label{exa:abstraction}
  Consider the \PRISM model \reffig{examples}(d).
  As seen in \refexa{prism}, its induced CTMC is the model in \reffig{examples}(a).
  With the partitioning of \refexa{prism} ($\astate_0 = \{\state_0\}$, $\astate_1 = \{\state_1, \state_2, \state_3, \state_4\}$, 
  and $\astate_2 = \{\state_5\}$), the model in \reffig{examples}(b) is an abstraction of the CTMC.
  Assume that the reward specification given denotes cumulative rewards.
  As we have already seen in \refexa{values}, the actual value for this model is 
  $\approx 2.70116$ while with the ECTMC abstraction we can bound the value to $\approx [1.82699,4.30277]$.
  If we split $\astate_1$, yielding the partitioning
  $\astate_0' = \{\state_0\}$, $\astate_1' = \{\state_1, \state_2\}$, $\astate_2' = \{\state_3, \state_4\}$, 
  and $\astate_3' = \{\state_5\}$, we obtain the refined ECTMC abstraction of \reffig{examples}(e).
  By doing so, the value bounds improve to $\approx [2.32375,3.21667]$.

  For comparison, we also provide the abstraction when using CTMDPs, cf.\ \reffig{examples}(b).
  When using a CTMDP, we have more transitions, because in this example we basically 
  have to represent the behaviour of each state by a distinct nondeterministic choice, 
  such that the abstraction is not much smaller than the original model.
  Indeed, as already discussed in the introduction, this problem is likely to occur 
  for models in which there is a large number of different rate values.
  In the example here, as seen from \refexa{values}, we obtain the same values 
  when using CTMDP or ECTMC abstractions.

  On the other hand, if we use abstract Markov chains (ACTMCs) \cite{KatoenKLW07} such as 
  in \reffig{examples}(f), the abstraction would be even smaller.
  However, it would also be less precise, as we would have to subsume the transitions with different domains of the successor 
  transition $\dom(\prismSucc(\state, \cdot))$.
  In this particular example, because of this constraint, we have a transition from $\astate_1$ to $\astate_2$ which states 
  that the transition probability is between $0$ and $1$, while for the ECTMC abstraction we do not 
  have intervals with a lower bound of $0$.
  The bounds we can obtain are $\approx [1.82699,4.38]$.
  Thus, the upper bound is larger than if using ECTMC abstractions.
\end{example}

With a given partitioning, we can apply \refalg{comp-imc} to obtain an
abstraction of the model. The algorithm computes an ECTMC to provide
an upper bound of the model value; a corresponding algorithm for the
lower bound can be defined likewise by minimising over the rewards of a given abstract state rather than maximising over them.
Indeed, we can use the same partitioning to obtain ECTMCs for the computation of both lower and upper bounds,
and thus compute abstractions for both directions at the same time.

The algorithm does some initializations, and afterwards, in line~\ref{lin:call-approx}, calls
\refalg{approx}. This algorithm descends into
the OBDD partitioning (lines~\ref{lin:descend_start} to \ref{lin:descend_end}), visiting each state of the model
explicitly. When a specific state $\state$ contained in an abstract
state $\astate$ is reached (lines~\ref{lin:visit-state-first} to
\ref{lin:visit-state-last}), we extend the abstracting model to take
into account the behaviour of this state. In lines
\ref{lin:widen-crew} and \ref{lin:widen-frew}, we extend the upper
bounds for the reward rates of $\astate$ such that they are at least
as high as those of $\state$. Notice that to compute the reward rates
of this state we use the original high-level PM, not the OBDD
representation. Then, in lines \ref{lin:widen-rates-begin} to
\ref{lin:widen-rates-end}, we handle the transition rates, thus to
include the rates of the concrete state. We again use the high-level
model, this time to compute the set of commands that are enabled in the current state
(line~\ref{lin:compute_commands}), the corresponding action
$\hat\alpha$ (line~\ref{lin:compute_action}, cf. \refdef{cmc-abs}), and the concrete successor states
with their corresponding rates (line~\ref{lin:compute_succ}). For each
of them, we use the function \fctcall{sAbs} to obtain the abstract state 
it belongs to. 
For all successor states, we add up the
rates to the same abstract state (line~\ref{lin:rate-addup}).
Then, starting from line~\ref{lin:widen-rates-assign},
we apply the actual widening of the rates.

Function \fctcall{sAbs} works as follows. Because we use a variable
order in which the variables encoding states are placed above the
variables for the abstract states, each state $\state\in\states$
induces a path in the OBDD $\partition$ which ends at the OBDD node
that represents the abstract state of $\state$. We follow the unique
path, given by the encoding of $s$, to the terminal node labelled with
1. This following yields the encoding of the abstract state of $s$. The running time
of \fctcall{sAbs} is therefore linear in the number of OBDD variables.

Let $n = |\states|$ be the number of concrete states of the model
under consideration, let $k$ be the total number of positive
transitions, and let $c$ be the number of OBDD variables.
\refalg{approx} visits each state and transition once. Because the block
number variables are placed at the bottom of the variable order,
accessing the block number of a state in  function \fctcall{sAbs}
has a running time of $\mathcal{O}(c)$. From this result, we
have that the overall complexity of \refalg{comp-imc} is
$\mathcal{O}((n + k) \cdot c)$.

\begin{algorithm}
  \caption{\label{alg:comp-imc}Compute ECTMC and reward structure from a given partitioning 
    $\apart$ with OBDD $\bdd_\apart = (\bddNodes, \bddNode_\apart, \bddHi, \bddLo, \bddNV)$ 
    of PM $\prismModel = (\prismVars, \prismInit, \prismCmds, \prismSucc, \prismCRew, \prismFRew)$.}
  \DontPrintSemicolon
  $\mathbf{global}\ \ivall, \ivalu, \crew, \frew$~(cf.~\refalg{approx}) \;
  $\ivall(\cdot,\cdot)(\cdot) \gets \mathrm{undefined}$ \;
  $\ivalu(\cdot,\cdot)(\cdot) \gets \mathrm{undefined}$ \;
  $\crew(\cdot) \gets \frew(\cdot) \gets -\infty$ \;
  $\astates \gets \{0,1,\ldots,|\apart|-1\}$ \;
  $\mathrm{approx}(\bddNode_\apart, 0)$ \label{lin:call-approx} \;
 \Return $\left((\astates, \ivall, \ivalu), (\crew, \frew)\right)$  \;
\end{algorithm}

\begin{algorithm}
  \newcommand{\level}{\mathrm{level}}
  \caption{\label{alg:approx}Procedure $\mathrm{approx}(\bddNode, \level)$.}
  \DontPrintSemicolon
  $\mathbf{global}\ \ivall, \ivalu, \crew, \frew$~(cf.~\refalg{comp-imc}) \;
  \lIf{$\bddNode = \bddZero$}{%
    \Return \;
  }
  \ElseIf{$\level < \mathrm{leafLevel}$}{
    \emph{// We are still at a variable level.}\label{lin:descend_start}\;
    $\bddVar = \mathrm{varAtLevel}(\level)$ \;
    \If{$\bddNode \neq \bddOne$ \textbf{\upshape and} $\bddVar = \bddNV(\bddNode)$}{%
      $\bddNode_l \gets \bddLo(\bddNode)$,
      $\bddNode_h \gets \bddHi(\bddNode)$ \;
    }
    \Else{%
      $\bddNode_l \gets \bddNode$,
      $\bddNode_h \gets \bddNode$ \;
    }
    \If{$\bddVar \in \avars$}{
      $\astate(\bddVar) \gets 0$,
      $\mathrm{approx}(\bddNode_l, \level+1)$ \;
      $\astate(\bddVar) \gets 1$,
      $\mathrm{approx}(\bddNode_h, \level+1)$ \;
    }
    \Else{
      $\state(\bddVar) \gets 0$,
      $\mathrm{approx}(\bddNode_l, \level+1)$ \;
      $\state(\bddVar) \gets 1$,
      $\mathrm{approx}(\bddNode_h, \level+1)$\label{lin:descend_end} \;
    }
  }
  \Else{ \emph{// We have traversed all variable levels.}  \label{lin:visit-state-first} \;
    $\crew(\astate) \gets \max(\crew(\astate), \prismCRew(\state))$ \label{lin:widen-crew} \;
    $\frew(\astate) \gets \max(\frew(\astate), \prismFRew(\state))$ \label{lin:widen-frew} \;
    \;
    $C \gets \dom(\prismSucc(\state,\cdot))$\label{lin:widen-rates-begin} \hfill\emph{// commands enabled in $s$} \label{lin:compute_commands} \;
    $\hat\alpha = \bigl\{ (c,\mathrm{sAbs}(\bddNode_\apart, s'))\,\big|\, c\in C\wedge s' = \prismNSucc(s,c)\bigr\}$ \label{lin:compute_action} \;
    $A \gets \prismSucc(\state)$ \label{lin:compute_succ} \;
    $\Lambda(\cdot) \gets 0$  \;
    \ForAll{$(\state', \lambda) \in A$} {
      $\astate' \gets \mathrm{sAbs}(\bddNode_\apart,
      \state')$ \label{lin:call-sabs} \;
        $\Lambda(\astate') \gets \Lambda(\astate') + \lambda$ \label{lin:rate-addup} \;
    }
    \ForAll{\upshape $\astate' \in \astates$}{ \label{lin:widen-rates-assign}
      $\ivall(\astate, \hat\alpha)(\astate') \gets \min\bigl(\ivall(\astate,\hat\alpha)(\astate'), \Lambda(\astate')\bigr)$ \;
      $\ivalu(\astate, \hat\alpha)(\astate') \gets \max\bigl(\ivalu(\astate,\hat\alpha)(\astate'), \Lambda(\astate')\bigr)$ \;
    } \label{lin:visit-state-last} \label{lin:widen-rates-end}
  }  
\end{algorithm}

In the discussion so far, we assumed that it is already clear how the
set of concrete states shall be divided into abstract states. We
might however come across models where this is not clear, or where the
results obtained from the abstraction are unsatisfactory. In these
cases, we have to apply \emph{refinement}, that is, split existing
abstract states into new ones. For other model types, such refinement
procedures already exist~\cite{KattenbeltKNP10,HermannsWZ08}. In the
analysis types considered before, schedulers sufficed which fix a
decision per state, and take neither the past history nor number of
steps before the state was entered into account. Then, depending on
the decisions of the scheduler per state, new predicates are
introduced to split the state space. In our case, such simple
schedulers are not sufficient to obtain extremal values, as has
already been shown for the simpler case of time-bounded
reachability~\cite{BaierHKH05}. Thus, it is not clear how to
introduce predicates to split the state space.

As a first heuristic, we do the following. We treat an OBDD
representation of a PM as a labelled transition system, in which the
commands play the role of the labels. We then use an existing
algorithm to symbolically compute (non-probabilistic) strong
bisimulations~\cite{wimmer-et-al-atva-2006}, but stop the algorithm
after a number of steps. This way, we obtain a partitioning in the
form of \refdef{bdd-partitioning}. As we will see later in
\refsec{case-studies}, although the method is not guaranteed to yield
a good abstraction, it can work well in practice.

The method discussed works for a very general class of PMs and
arbitrary state partitionings. However, because it is based on
explicitly visiting each concrete state at least once, it may take
much time to perform for large models.  To tackle this problem, a
parallel implementation of the technique is possible.  Given a
computing system with a number of processors, one can symbolically
divide the states of the model, such that each processor works on a
different part of the OBDD representing the state space.  Each
processor can then process the model part it is assigned to.  The only
point of interaction is the widening of the rates and reward rates of
the abstract model.  On a shared memory architecture, one could use
different semaphores for the reward rates and successor transitions of
each abstract state to avoid delays.  Without shared memory, the
processors can compute partial abstractions separately, which are
merged after the computations are finished.  This technique is faster,
but has the disadvantage of having to store several (partial) copies
of the abstraction.  If the state space is divided such that all
states of an abstract state are assigned to a single processor, no
locking is needed, and the overhead is reduced.

As an alternative to parallelisation, it should also be possible to
use optimisation methods over variants of
BDDs~\cite{LaiPV94,Knuth09,OssowskiB06,WanCM11} to compute the rate
and reward intervals symbolically rather than rely on explicit
enumeration of all possible variable assignments.

\section{Case Studies}
\label{sec:case-studies}
\noindent To show the practicality of the method, we applied it on two case
studies from classical performance and
dependability engineering \cite{HaverkortHK00,ClothH05}.
We implemented the techniques of \refalg{opt-reach}, and \refalg{comp-imc}.
To represent the ECTMCs, we used a sparse-matrix-like data structure.

Where possible, we compared the results to \PRISM.
\PRISM always starts by building an MTBDD representation of the model under consideration.
The subsequent analysis is then performed using \emph{value iteration} in the CTMC semantics similarly to \refalg{opt-reach}.
The data structure used here is either an MTBDD, a sparse matrix, or a \emph{hybrid} structure~\cite{KwiatkowskaNP04}.
In the latter, values for the model states are stored explicitly, but parts of the transition structure are stored implicitly.

For all experiments, we used a Quad-Core AMD Opteron\texttrademark{} Processor 8356 (of which we only used one core)
with 2300~MHz, and 64~GB of main memory.

  \setlength{\tabcolsep}{0.6mm}
\begin{table*}
  \centering
  \caption{\PRISM results for the number of repairs in the workstation cluster until $\timeb=500$}
  \label{tab:cluster_detailed_prism}
\begin{tabular}{|rrr|rr|rr|rr|r|}
  \hline
    && & \multicolumn{2}{c|}{sparse engine} & \multicolumn{2}{c|}{hybrid engine} & \multicolumn{2}{c|}{symbolic engine} & \\
  $N$ & \multicolumn{1}{c}{$|S|$} & \multicolumn{1}{c|}{$|R|$} & Time & Memory & Time & Memory & Time & Memory & \multicolumn{1}{c|}{Result} \\
  \hline
  32 & $ 3.87 \cdot 10^{4} $ & $ 1.86 \cdot 10^{5} $ & 9.29 & 36.67 & 14.90 & 37.48 & 13\,791.20 & 184.49 & 64.17635 \\
  64 & $ 1.51 \cdot 10^{5} $ & $ 7.33 \cdot 10^{5} $ & 62.11 & 42.92 & 88.93 & 41.69 & \multicolumn{2}{c|}{\timelimit} & 127.98101 \\
  128 & $ 5.97 \cdot 10^{5} $ & $ 2.91 \cdot 10^{6} $ & 380.51 & 60.18 & 585.55 & 54.48 & \multicolumn{2}{c|}{\timelimit} & 255.48297 \\
  256 & $ 2.37 \cdot 10^{6} $ & $ 1.16 \cdot 10^{7} $ & 3\,182.73 & 141.71 & 4\,737.73 & 98.27 & \multicolumn{2}{c|}{\timelimit} & 509.58417 \\
  512 & $ 9.47 \cdot 10^{6} $ & $ 4.62 \cdot 10^{7} $ & 10\,540.54 & 817.39 & 14\,965.74 & 284.66 & \multicolumn{2}{c|}{\timelimit} & 896.80612 \\
  1\,024 & $ 3.78 \cdot 10^{7} $ & $ 1.85 \cdot 10^{8} $ & 13\,242.08 & 3\,154.91 & 25\,513.31 & 1\,014.79 & \multicolumn{2}{c|}{\timelimit} & 905.19921 \\
  2\,048 & $ 1.51 \cdot 10^{8} $ & $ 7.39 \cdot 10^{8} $ & \multicolumn{2}{c|}{\timelimit} & \multicolumn{2}{c|}{\timelimit} & \multicolumn{2}{c|}{\timelimit} & ?? \\
  4\,096 & $ 6.04 \cdot 10^{8} $ & $ 2.95 \cdot 10^{9} $ & \multicolumn{2}{c|}{\memlimit} & \multicolumn{2}{c|}{\timelimit} & \multicolumn{2}{c|}{\timelimit} & ?? \\
  8\,192 & $ 2.42 \cdot 10^{9} $ & $ 1.18 \cdot 10^{10} $ & \multicolumn{2}{c|}{\memlimit} & \multicolumn{2}{c|}{\memlimit} & \multicolumn{2}{c|}{\timelimit} & ?? \\
  16\,384 & $ 9.66 \cdot 10^{9} $ & $ 4.72 \cdot 10^{10} $ & \multicolumn{2}{c|}{\memlimit} & \multicolumn{2}{c|}{\memlimit} & \multicolumn{2}{c|}{\timelimit} & ?? \\
  \hline
\end{tabular}
\end{table*}

\begin{table}[tb]
  \centering
  \caption{\label{tab:cluster}Number of repairs in the workstation cluster until time $\timeb=500$}
\begin{tabular}{|r|rrrc|}
\hline
     & \multicolumn{4}{c|}{ECTMC Results} \\
 $N$ & $|\apart|$ & Time & Memory & Interval \\
\hline
32 & 19\,420 & 107.15 & 90.03 & $[ 64.176, 64.199]$ \\
64 & 19\,420 & 109.43 & 86.28 & $[ 127.980, 128.490]$ \\
128 & 19\,420 & 115.93 & 89.54 & $[ 255.455, 259.797]$ \\
256 & 19\,420 & 132.89 & 94.58 & $[ 509.000, 580.485]$ \\
512  & 19\,420 & 181.99 & 91.87 & $[ 869.749, 900.052]$ \\
1\,024 & 19\,420 & 412.43 & 107.22 & $[ 905.018, 905.200]$ \\
2\,048 & 19\,420 & 1\,335.54 & 103.31 & $[ 905.766, 905.767]$ \\
4\,096 & 19\,420 & 5\,298.29 & 104.89 & $[ 905.955, 905.955]$ \\
8\,192 & 19\,420 & 28\,361.36 & 132.48 & $[ 906.040, 906.040]$ \\
16\,384 & 19\,420 & 147\,691.30 & 139.56 & $[ 906.084, 906.084]$ \\
\hline
\end{tabular}
\end{table}
%
We consider a fault-tolerant workstation cluster~\cite{HaverkortHK00}. It consists of two
sub-clusters, which, in turn, contain $N$ workstations connected via a
central switch. The two switches are connected via a backbone. Each
component of the system can break down, and is then fixed by a single
repair unit responsible for the entire system.

We are interested in the expected number of repairs until a time bound of $\timeb = 500$.
This property can be expressed using cumulative
rewards. For $N$ up to $512$, the model has been successfully
analysed before using \PRISM
\footnote{\url{http://www.prismmodelchecker.org/casestudies/cluster.php\#mc},
Property \texttt{R\{"num\_repairs"\}=?[ C<=T ]}. }. While the
existing analysis methods worked well for model instantiations up to
this $N$, and somewhat above, the techniques do not work well anymore
for a very large number of workstations. Constructing the model using
MTBDDs seems not to be problematic, but the subsequent analyses cannot
be performed successfully. The sparse-matrix and the hybrid method
fail at some point, because they rely on an explicit representation of
the state space, and thus run out of memory. Also, the MTBDD-based
value iteration fails at some point, and works rather slowly. The
reason for this failure is probably that, during the value iteration, a large number of
different non-terminal nodes appear, which make the MTBDD complex, and
thus large and slow to operate on. Detailed information about the
performance of \PRISM on this case study is given in \reftab{cluster_detailed_prism}.

By $|S|$, and $|R|$, we give the approximate number of states, and
transitions, resp., of the original CTMC model. For each of the three
\PRISM engines, we give the running time (columns labelled with Time), and 
memory consumption (columns labelled with Memory) for computing
the expected rewards. An entry of \timelimit{} means that \PRISM did 
not terminate within 160,000 seconds, while an entry of \memlimit{} indicates that 
more than 60\,GB of memory are required to complete the analysis.

In \reftab{cluster}, we apply the method developed in this
paper on several instantiations of the number of workstations $N$.  
The results we obtained by our method are given in ECTMC Results.
Besides the running time and memory consumption, we give in the column
titled $|\apart|$ the number of abstract states we used for the corresponding analysis.
The column labelled Interval gives the lower and upper bounds of the actual value of the expected reward.

As we see from the time and memory usage, for smaller models, it is
advantageous to use an explicit-state method as implemented in \PRISM,
because of the additional overhead our method introduces. As
instantiations become larger, using the method of this paper becomes
worthwhile. While we do not always get precise bounds for all
analyses performed with this number of abstract states, we always were
able to compute the order of magnitude. Interestingly, the value
bounds get tighter with an increasing number of model states.

\begin{figure*}[tb]
  \centering
  \begin{tabular}{cc}
  \includegraphics[width=0.4\textwidth]{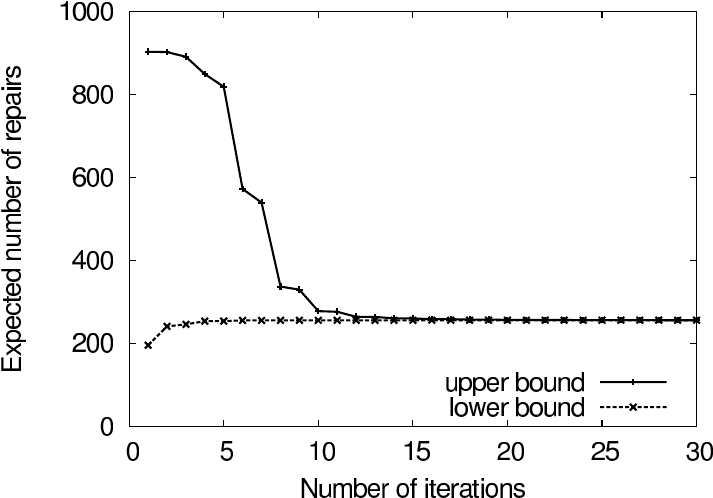} & \includegraphics[width=0.4\textwidth]{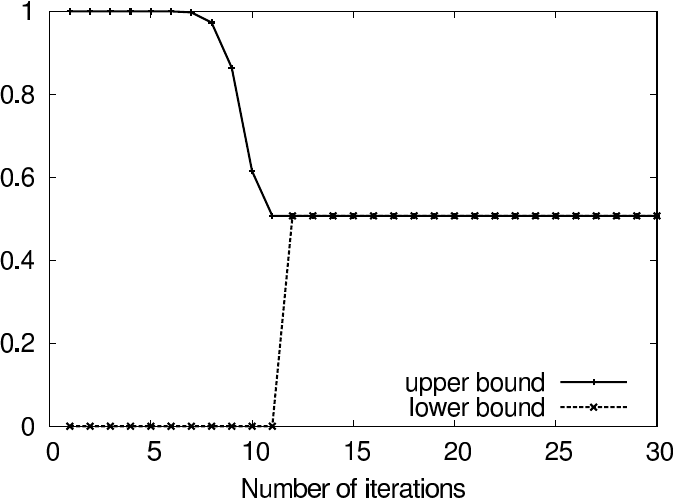} \\
  \footnotesize (a) Workstation cluster with $N=128$. & \footnotesize (b) Google file system with $M=256$.
  \end{tabular}
  \caption{Quality of the ECTMC approximation for different numbers of bisimulation iterations.}
  \label{fig:quality}
\end{figure*}

As discussed in \refsec{algorithms}, we apply a heuristic refinement algorithm based on bisimulations for labelled transition systems.
We use the symbolic algorithm \cite{wimmer-et-al-atva-2006} for computing (non-stochastic) strong bisimulations to obtain a suitable state partitioning.
We abort its fix-point iteration prematurely after a user-specified number $n$ of iterations.
In \reffig{quality}, we show how the quality of the
approximation evolves depending on $n$. The behaviour of the cluster
case study is shown on the left. One can observe that the
width of the computed interval converges quickly to the actual value
when increasing the number of iterations.

Notably, if we use the same number of refinement steps, for all model instantiations
considered, $|\apart|$ stayed constant, although the number of model
states $|\states|$ was different for each instantiation (cf. \reftab{cluster}).

\begin{table*}[tb]
  \centering
  \caption{Detailed experimental results for the workstation cluster ($N=2048$, $\timeb=500$) using the ECTMC abstraction}
  \label{tab:cluster_detailed_cmc}
\begin{tabular}{|rr|rrrrr|r|c|}
  \hline
             &              & \multicolumn{5}{c|}{Running Times}         &             &  \\
  Iterations & $|\apart|$ & \NUSMV & Refinement & ECTMC & Value Iter. & Total & Memory & Interval \\
  \hline
  5 & 2440 & 64.75 & 0.93 & 991.16 & 13.45 & 1070.34 & 56.13 & $[902.092, 905.771]$ \\
  10 & 9216 & 68.46 & 6.96 & 1135.57 & 41.68 & 1252.79 & 70.11 & $[905.739, 905.767]$ \\
  15 & 19420 & 65.25 & 18.26 & 1029.23 & 116.89 & 1229.78 & 103.31 & $[905.766, 905.767]$ \\
  20 & 34596 & 73.93 & 43.71 & 1150.70 & 213.72 & 1482.29 & 134.62 & $[905.767, 905.767]$ \\
  25 & 52600 & 69.82 & 78.88 & 1070.13 & 321.43 & 1540.49 & 180.85 & $[905.767, 905.767]$ \\
  30 & 76176 & 66.43 & 126.85 & 1052.87 & 466.31 & 1712.76 & 227.80 & $[905.767, 905.767]$ \\
  \hline
\end{tabular}
\end{table*}
\reftab{cluster_detailed_cmc} contains more detailed running times for
the cluster benchmark with $N=2048$ workstations using the ECTMC
abstraction for different numbers of bisimulation iterations, which
are given in the first column. The second column contains the number
of abstract states. The running times in seconds are given separately
for the different main operations. The running times include the call to \NUSMV to generate the
OBDDs for the underlying transition system (col.~3), the given number
of bisimulation iterations (col.~4), the construction of the ECTMC
from the partitioning (col.~5), the value iteration to compute the
reward interval (col.~6), and finally the total computation time
(col.~7). The last two columns contain the memory consumption in
Megabytes, and the computed reward interval.

We additionally consider a replicated file system as used as part of
the Google search engine~\cite{ClothH05,BaierHHHK12}. Originally, the model was given as a
generalised stochastic Petri net, but was transformed to a PM for the
analysis.

\begin{table*}
  \centering
  \caption{\PRISM results for the Google file system}
  \label{tab:google_detailed_prism}
\begin{tabular}{|rrr|rr|rr|rr|r|}
  \hline
    &&  & \multicolumn{2}{c|}{sparse engine} & \multicolumn{2}{c|}{hybrid engine} & \multicolumn{2}{c|}{symbolic engine} & \\
  $N$ & \multicolumn{1}{c}{$|S|$} & \multicolumn{1}{c|}{$|R|$} & Time & Memory & Time & Memory & Time & Memory & \multicolumn{1}{c|}{Result} \\
  \hline
  32  & $ 6.15 \cdot 10^{3} $ & $ 4.03 \cdot 10^{4} $ & 1.60 & 607.42 & 1.68 & 607.78 & 67.95 & 624.01 & 0.000000 \\
  64  & $ 2.46 \cdot 10^{4} $ & $ 1.66 \cdot 10^{5} $ & 34.72 & 614.24 & 71.04 & 616.84 & 66319.89 & 791.23 & 0.507119\\
  128 & $ 9.83 \cdot 10^{4} $ & $ 6.77 \cdot 10^{5} $ & 464.25 & 624.81 & 1\,890.41 & 627.23 & \multicolumn{2}{c|}{\timelimit} & 0.507119 \\
  256 & $ 3.93 \cdot 10^{5} $ & $ 2.74 \cdot 10^{6} $ & 3\,083.24 & 683.27 & 26\,132.52 & 701.11 & \multicolumn{2}{c|}{\timelimit} & 0.507119 \\
  512 & $ 1.57 \cdot 10^{6} $ & $ 1.11 \cdot 10^{7} $ & 41\,901.82 & 938.10 & \multicolumn{2}{c|}{\timelimit} & \multicolumn{2}{c|}{\timelimit} & 0.507119 \\
  1\,024 & $ 6.29 \cdot 10^{6} $ & $ 4.44 \cdot 10^{7} $ & \multicolumn{2}{c|}{\timelimit} & \multicolumn{2}{c|}{\timelimit} & \multicolumn{2}{c|}{\timelimit} & ?? \\
  2\,048 & $ 2.52 \cdot 10^{7} $ & $ 1.78 \cdot 10^{8} $ & \multicolumn{2}{c|}{\timelimit} & \multicolumn{2}{c|}{\timelimit} & \multicolumn{2}{c|}{\timelimit} & ?? \\
  \hline
\end{tabular}
\end{table*}

\begin{table}[tb]
  \centering
  \caption{\label{tab:google}Google file system with property 12 and $\timeb=60$, $N=100\,000$, $C=5000$}
\begin{tabular}{|r|rrrc|}
\hline
     & \multicolumn{4}{c|}{ECTMC Results} \\
 $M$ & $|\apart|$ & Time & Mem. & Interval \\
\hline
32 & 6\,138 & 30.28 & 72.28 & $[ 0.0000, 0.0000]$ \\
64 & 18\,042 & 251.86 & 86.53 & $[ 0.5071, 0.5071]$ \\
128 & 29\,515 & 822.08 & 142.23 & $[ 0.5071, 0.5071]$ \\
256 & 29\,515 & 1\,531.46 & 147.66 & $[ 0.5071, 0.5071]$ \\
512 & 29\,515 & 2\,951.37 & 129.72 & $[ 0.5071, 0.5071]$ \\
1\,024 & 29\,515 & 6\,776.42 & 128.40 & $[ 0.5071, 0.5071]$ \\
2\,048 & 29\,515 & 14\,957.94 & 137.47 & $[ 0.5071, 0.5071]$ \\
\hline
\end{tabular}
\end{table}

Files are divided into \emph{chunks} of equal size.  Several copies of
each chunk reside at several \emph{chunk servers}.  There is a single
\emph{master} server which knows the location of the chunk copies.  If
a user of the file system wants to access a certain chunk of a file,
it asks the master for the location.  Data transfer then takes place
directly between a chunk server and the user.  The model describes the
life cycle of a single chunk, but accounts for the load caused by the
other chunks.

The model features three parameters: $M$ is the number of chunk
servers, with $C$ we denote the number of chunks a chunk server may
store, and the total number of chunks is $N$.

We consider the minimal probability over all states in which severe
hardware problems have occurred (the master server is down, and more than
three quarters of the chunk servers are down), that within time
$\timeb$ a state will be reached in which a guaranteed
quality-of-service level (all three chunk copies are present, and the
master server is available) holds.  This is a bounded-reachability
property, and thus based on final rewards.

We fix $C=5000$, $N=100\,000$, and $\timeb=60$; and we provide results for
several $M$ in \reftab{google}. In the analyses with \PRISM (see \reftab{google_detailed_prism}), we used
an improved OBDD variable order, such that the performance results are
better than in~\cite{BaierHHHK12}. In contrast
to the previous case study, \PRISM's sparse matrix engine was faster, and
did not use more memory compared to the hybrid engine. The symbolic 
engine was again the slowest. The MTBDD representation of the model 
requires more memory per concrete state compared to the previous case 
study. We assume that this requirement is because the number of different rates 
occurring is much higher, and because some of the rates are obtained by multiplying
state variables, thus leading to a more complex MTBDD structure.
Notice that in this model, from a certain value of $M$ onward, the
probability discussed is almost the same.

We give detailed information on the instance with $M=128$ of the Google
file system in \reftab{google_detailed_cmc} and \reffig{quality} (right-hand side)
for different numbers $n$ of bisimulation iterations.
We can again observe that the computed interval for the bounded-reachability
property quickly converges to the actual probability with increasing $n$.

\begin{table*}[tb]
  \centering
  \caption{Detailed experimental results for the Google file system ($M=128$, $\timeb=60$) using the ECTMC abstraction}
  \label{tab:google_detailed_cmc}
\begin{tabular}{|rr|rrrrr|r|c|}
  \hline
               &                & \multicolumn{5}{c|}{Running Times} & &  \\
  Iterations & $ |\apart|$ & \NUSMV & Refinement & ECTMC & Value Iter. & Total & Memory & Interval \\
  \hline
  5 & 3204 & 0.86 & 3.19 & 1.80 & 70.42 & 76.98 & 65.00 & $[0.000000, 0.999999]$ \\
  10 & 15564 & 0.81 & 15.44 & 2.01 & 426.32 & 445.42 & 92.08 & $[0.000000, 0.614823]$ \\
  15 & 29515 & 0.83 & 41.24 & 2.20 & 716.67 & 761.89 & 142.23 & $[0.507119, 0.507119]$ \\
  20 & 42725 & 0.83 & 82.54 & 2.41 & 1040.74 & 1127.53 & 177.59 & $[0.507119, 0.507119]$ \\
  25 & 57472 & 0.79 & 137.50 & 2.57 & 1349.73 & 1491.65 & 249.21 & $[0.507119, 0.507119]$ \\
  30 & 69932 & 0.82 & 191.83 & 2.62 & 1652.66 & 1849.04 & 279.13 & $[0.507119, 0.507119]$ \\
  \hline
\end{tabular}
\end{table*}

From Table~\ref{tab:cluster}, we see that the obtained reward converges to a fixed value with increasing model parameter $N$.
Therefore, if one is interested in the value obtained if the model parameters go towards infinity, methods concerned with the limiting behaviour~\cite{BortolussiH12,BortolussiH13,HenzingerMMW10,AngiusHW13} might be more appropriate.
As discussed in the introduction, our method targets at finding conservative bounds for the model under consideration.
The asymptotic methods we are aware of do not provide such guaranteed bounds in the general case.
Also, the models which we used are specified in a very general guarded commands language, and cannot be adequately modelled in a restricted input language, as usually required as the input format of asymptotic analyses.
Our method provides correct results for all choices of the parameters, whereas for smaller or medium model sizes the results obtained by asymptotic analyses would be very far off, even if it would be possible to perform such an analysis for increasing parameter sizes.
Many scalable models (e.g. many of the ones from the homepage of the probabilistic model checker \PRISM) behave in a similar way.

\section{Conclusion}
\label{sec:conclusion}
\noindent We developed an efficient method to compute extremal values of CTMDPs over HR schedulers.
It can be used to safely bound quantities of interest of CTMCs, by abstracting them into a special class of CTMDPs, and then applying this method.
Experimental results have shown that the approach works well in practice.

There are a number of possible future works.
The current refinement technique surely does not yield an optimal partitioning of the state space in all cases.
We thus want to see how the scheduler we implicitly obtain by \refalg{opt-reach} can be used to refine the model.
The abstraction technique could also be extended to other property classes and models.
For instance, models already involving nondeterminism could be abstracted and approximated using Markov games~\cite{BrazdilFKKK09,RabeS10,RabeS11}.
It would also be interesting to see how a parallelised or symbolic abstraction method sketched at the end of \refsec{algorithms} performs against the one currently implemented.
Using a three-valued logic~\cite{Klink10}, the technique could also be integrated into an existing probabilistic (CSL) model checker.

\balance

\clearpage

\section*{Appendix}

\label{appendices}

\subsection*{Proof of \refpro{algo-correctness}}
\label{apx:algo-correctness}
%


\begin{definition}
  \label{def:transprobsksp}
  Let $\dmodel = (\states, \pmat)$ be a DTMC.
  For $\state_0, \state_k \in \states$, and $k \in \nats$, we define $\tprob^\dmodel(\state_0, k, \state_k) = \prob(\stopro^{\dmodel,\state_0}_k = \state_k)$.
\end{definition}
\refdef{transprobsksp} specifies the transient probability to be in state $\state_k$ at step $k$ when having started in state $\state_0$.
\begin{corollary}
\label{cor:tprob-matrix}
Notice that, for $\dmodel = (\states, \pmat)$, $k \in \nats$, and $\state_0, \state_k \in \states$, it holds that
\begin{equation*}
\begin{split}
& \tprob^\dmodel(\state_0, k, \state_k) \\
= {} & \pmat^k(\state_0, \state) \\
\defeq & \sum_{\state_1 \in \states} \pmat(\state_0, \state_1) \cdot \sum_{\state_2 \in \states} \pmat(\state_1, \state_2) \cdot \sum_{\state_3 \in \states} \pmat(\state_2, \state_3) \\
& \qquad\cdot\,\dotsm\,\cdot \sum_{\state_{k-1}\in\states} \pmat(\state_{k-2}, \state_{k-1}) \cdot \pmat(\state_{k-1}, \state_k),
\end{split}
\end{equation*}
i.\,e., the transient probability in a DTMC can be expressed using matrix multiplications.
\end{corollary}

We extend the definition of values to discrete-time models, which will be used in the further parts of the proof.
\begin{definition}
\label{def:disc-value}
  Let $\dmodel = (\states, \pmat)$ be a DTMC, let $\rew = (\crew, \frew)$ be a reward structure, and let $\urate, \timeb \geq 0$.
  We define
  \begin{equation*}
  \begin{split}
    & \mvalue(\dmodel, \state_0, \rew, \timeb, \urate) \\
    \defeq & \sum_{i=0}^\infty \left(\phi_{\urate \timeb}(i) \sum_{\state_i \in \states} \tprob^{\dmodel}(\state_0, i, \state_i) \frew(\state_i)\right. \\
    & + \left.\psi_{\urate \timeb}(i) \sum_{\state_i \in \states} \tprob^{\dmodel}(\state_0, i, \state_i)
  \frac{\crew(\state_i)}{\urate}\right) .
  \end{split}
\end{equation*}
\end{definition}

We define a type of schedulers which are simpler than the HRs of \refdef{sched-hrs}, and at the same time generalise the CD of \refdef{sched-cd}.
\begin{definition}
  \label{def:sched-cr}
  A \emph{time-abstract, history-abstract, counting, randomised scheduler (CR)} for a DTMDP $\dmodel = (\states, \acts, \pmat)$ or a CTMDP $\cmodel = (\states, \acts, \rmat)$ is a function $\sched\colon (\states \times \nats) \to \distrs(\acts)$
  such that, for all $\state \in \states$, and $n \in \nats$, if $\sched(\state, n)(\act) > 0$, then $\act \in \acts(\state)$.
  With  $\schedscr$, we denote the set of all CRs.
\end{definition}

\begin{definition}
  \label{def:induced}
  Assume we are given a CTMDP $\cmodel = (\states, \acts, \rmat)$, and a CR
  $\sched\colon (\states \times \nats) \to \distrs(\acts)$. We define the \emph{induced CTMC} as $\cmodel_\sched \defeq (\states', \rmat')$ with
  \begin{itemize}
  \item $\states' \defeq \states \times \nats$,
  \item $\rmat'((\state, n), (\state', n+1)) \defeq \sum_{\act \in \acts} \sched(\state, n)(\act) \cdot \rmat(\state, \act, \state')$ for $\state, \state' \in \states$ and $n \in \nats$, and $\rmat'(\cdot) \defeq 0$ else.
  \end{itemize}
  Let $\stopro^{\cmodel_\sched,\state_0}\colon (\samspace{\cmodel_\sched} \times \reals_{\geq 0}) \to (\states \times \nats)$ be the stochastic process of the CTMC $\cmodel_\sched$, and let $f\colon (\states \times \nats) \to \states$ with $f(\state, n) = \state$.
  Induced DTMCs of DTMDPs are defined accordingly.
  The \emph{induced stochastic process} $\stopro^{\cmodel,\sched,\state_0}\colon (\samspace{\cmodel_\sched} \times \reals_{\geq 0}) \to \states$ of $\cmodel$ and $\sched$ starting in $\state_0 \in \states$ is then defined as $\stopro^{\cmodel,\sched,\state_0}_t = f \circ \stopro^{\cmodel_\sched,(\state_0,0)}_t$ for $t \in \reals_{\geq 0}$.
  Definitions for DTMDPs are likewise using $\pmat$ instead of $\rmat$.
\end{definition}

We extend the notation of transient probabilities to scheduled nondeterministic models.
It is known that, for DTMDPs, CR schedulers are as powerful as HR schedulers.
\begin{definition}
  For a DTMDP $\dmodel = (\states, \acts, \pmat)$, and a scheduler $\sched \in \schedshr \cup \schedscd \cup \schedscr$,
  we define $\tprob^{\dmodel,\sched}(\state_0, k, \state_k) = \prob(\stopro^{\dmodel,\sched,\state_0}_k = \state_k)$ for all $k \in \nats$ and $\state_0, \state_k \in \states$.
\end{definition}

\begin{lemma}
  \label{lem:cr-suffices-dt}
  Consider a DTMDP $\dmodel = (\states, \acts, \pmat)$, and a HR $\schedhr$.
  Then there is a CR $\schedcr$ such that, for all $\state_0, \state_n \in \states$, and $n \in \nats$,
  we have $\tprob^{\dmodel,\schedhr}(\state_0, n, \state_n) = \tprob^{\dmodel,{\schedcr}}(\state_0, n, \state_n)$.
\end{lemma}
\begin{proof}
The proof is given in \cite{Strauch66} and \cite[Theorem 5.5.1]{Puterman94}, where CR are denoted as MR (Markov randomised) policies.
\end{proof}

\begin{definition}
  For a CTMC $\cmodel = (\states, \rmat)$, we let $\emb(\cmodel) \defeq (\states, \pmat)$ denote the DTMC
  such that, for all $\state, \state' \in \states$, we have $\pmat(\state, \state') \defeq \rmat(\state, \state')/\urate(\cmodel)$.
\end{definition}

The following lemma states how values of CTMCs can be computed using the embedded discrete-time model.
\begin{lemma}
  \label{lem:disc-rew}
  Let $\cmodel = (\states, \rmat)$ be a CTMC with a reward structure $\rew = (\crew, \frew)$.
  Let $\urate = \urate(\cmodel)$.
  Then, for $\timeb \geq 0$, and all $\state_0 \in \states$, the following holds.
  \begin{equation*}
    \mvalue(\cmodel, \state_0, \rew, \timeb) = \mvalue(\emb(\cmodel), \state_0, \rew, \timeb, \urate)
  \end{equation*}
\end{lemma}

\begin{proof}
By \refdef{value}, for $\state_0 \in \states$, it holds that
\begin{equation*}
\mvalue(\cmodel, \state_0, \rew, \timeb) = \underbrace{\expect \left[\int_0^\timeb \! \crew(\stopro_u^{\cmodel,\state_0}) \, \diff u \right]}_{\mathrm{accumulated}} + \underbrace{\expect \left[\frew(\stopro_\timeb^{\cmodel,\state_0}) \right]}_{\mathrm{final}} .
\end{equation*}
Thus, we can divide $\mvalue(\cmodel, \state_0, \rew, \timeb)$ into a sum of $\mathrm{accumulated}$ and $\mathrm{final}$.
We have
{\allowdisplaybreaks\begin{align*}
\mathrm{final} = {} & \expect \left[\frew(\stopro_\timeb^{\cmodel,\state_0}) \right] \\*
= {} & \sum_{\state \in \states} \prob(\stopro_\timeb^{\cmodel,\state_0} = \state) \frew(\state) \\*
= {} & \sum_{\state \in \states} \left(\sum_{i=0}^\infty \tprob^{\emb(\cmodel)}(\state_0, i, \state) \phi_{\urate\timeb}(i)\right) \frew(\state) \\*
= {} & \sum_{i=0}^\infty \phi_{\urate \timeb}(i) \sum_{\state \in \states} \tprob^{\emb(\cmodel)}(\state_0, i, \state) \frew(\state) . \\
\intertext{Further, Kwiatkowska et al.~\cite[Theorem 1]{KwiatkowskaNP06} have shown that}
\mathrm{accumulated} = {} & \sum_{i=0}^\infty \psi_{\urate \timeb}(i) \sum_{\state \in \states} \tprob^{\emb(\cmodel)}(\state_0, i, \state) \frac{\crew(\state)}{\urate} . \\
\intertext{Thus,}
  \mvalue(\cmodel, \state_0, \rew, \timeb)  = {} & \mathrm{accumulated} + \mathrm{final}\\*
 = {} & \sum_{i=0}^\infty \left(\phi_{\urate \timeb}(i) \sum_{\state \in \states} \tprob^{\emb(\cmodel)}(\state_0, i, \state) \frew(\state)\right. \\*
  & + \left.\psi_{\urate \timeb}(i) \sum_{\state \in \states} \tprob^{\emb(\cmodel)}(\state_0, i, \state)
  \frac{\crew(\state)}{\urate}\right) \\*
 = {} & \mvalue(\emb(\cmodel), \state_0, \rew, \timeb, \urate) .
\end{align*}}
\end{proof}

We can now show that the restricted class CR suffices to obtain optimal values.
\begin{lemma}
  \label{lem:cr-sameval-hrs}
  Given a CTMDP $\cmodel = (\states, \acts, \pmat)$, and $\schedhr \in \schedshr$,
  there is $\schedcr \in \schedscr$ such that
  $\mvalue(\stopro^{\cmodel,\schedhr,\state_0}, \rew, \timeb) = \mvalue(\stopro^{\cmodel,\schedcr,\state_0}, \rew, \timeb)$.
  Further, for all $\sigma_\mathit{cr}' \in \schedscr$, we can find $\sigma_\mathit{hr}' \in \schedshr$ such that
  $\mvalue(\stopro^{\cmodel,\sigma_\mathit{cr}',\state_0}, \rew, \timeb) = \mvalue(\stopro^{\cmodel,\sigma_\mathit{hr}',\state_0}, \rew, \timeb)$.
\end{lemma}

\begin{proof}
  Consider a CTMDP $\cmodel = (\states, \acts, \pmat)$, and $\schedhr \in \schedshr$.
  By \reflem{cr-suffices-dt}, we can find a scheduler $\schedcr \in \schedscr$ such that
  \begin{equation}\label{eqn:schedeqtrans}
    \tprob^{\emb(\cmodel),{\schedhr}} = \tprob^{\emb(\cmodel),\schedcr} .
  \end{equation}
  Define
  \begin{itemize}
  \item $\rew^\mathit{hr} \defeq (\crew^{\mathit{hr}}, \frew^{\mathit{hr}})$ with
  \item $\crew^{\mathit{hr}}(\mpath, \state) \defeq \crew(\state)$, and
  \item $\frew^{\mathit{cr}}(\mpath, \state) \defeq \frew(\state)$;
  \end{itemize}
  and let
  \begin{itemize}
  \item $\rew^{\mathit{cr}} \defeq (\crew^{\mathit{cr}}, \frew^{\mathit{cr}})$ with
  \item $\crew^{\mathit{cr}}(\state, n) \defeq \crew(\state)$, and
  \item $\frew^{\mathit{cr}}(\state, n) \defeq \frew(\state)$
  \end{itemize}
  for $\mpath \in (\states \times \acts)^*$, $n \in \nats$, and $\state \in \states$.
  Then for all $\state_0 \in \states$, we have
  \begin{align*}
    & \expect [ \frew(\stopro^{\cmodel,\schedhr,\state_0}_\timeb) ] \\
   {} = {}& \expect [ \frew(f(\stopro^{\cmodel_{\schedhr},\state_0}_\timeb)) ] \\
   {} = {}& \sum_{\state \in \states} \prob(f(\stopro^{\cmodel_{\schedhr},\state_0}_\timeb) = \state) \frew(\state) \\
   {} = {}& \sum_{\state \in \states} \prob(\exists \mpath \in (\states \times \acts)^* .\ \stopro^{\cmodel_{\schedhr},\state_0}_\timeb = (\mpath, \state)) \frew(\state) \\
   {} = {}& \sum_{\substack{\state \in \states, \\ \mpath \in (\states \times \acts)^*}} \prob(\stopro^{\cmodel_{\schedhr},\state_0}_\timeb = (\mpath, \state)) \frew(\state) \\
   {} = {}& \sum_{\substack{\state \in \states, \\ \mpath \in (\states \times \acts)^*}} \prob(\stopro^{\cmodel_{\schedhr},\state_0}_\timeb = (\mpath, \state)) \frew^\mathit{hr}(\mpath, \state) \\
   {} = {}& \expect [ \frew^\mathit{hr}(\stopro^{\cmodel_\schedhr,\state_0}_\timeb) ], \\
  \intertext{similarly}
    & \expect \left[\int_0^\timeb \! \crew(\stopro_u^{\cmodel,\schedhr,\state_0}) \, \diff u \right] \\
    {}={} & \int_0^\timeb \! \expect \left[ \crew(\stopro_u^{\cmodel,\schedhr,\state_0}) \right] \, \diff u \\
    {}={} & \int_0^\timeb \! \expect \left[ \crew^\mathit{hr}(\stopro_u^{\cmodel_{\schedhr},\state_0}) \right] \, \diff u \\
    {}={} & \expect \left[ \int_0^\timeb \! \crew^\mathit{hr}(\stopro_u^{\cmodel_{\schedhr},\state_0})  \, \diff u \right],
  \end{align*}
  and in turn
  \begin{align}\label{eqn:hrsrewrewpr}
    \mvalue(\stopro^{\cmodel,\schedhr,\state_0}, \rew, \timeb) &= \mvalue(\stopro^{\cmodel_{\schedhr},\state_0}, \rew^\mathit{hr}, \timeb) .
  \intertext{In the same way, using $n \in \nats$ instead of $\mpath \in (\states \times \acts)^*$, one can show}
  \label{eqn:crrewrewpr}
    \mvalue(\stopro^{\cmodel,\schedcr,\state_0}, \rew, \timeb) &=  \mvalue(\stopro^{\cmodel_\schedcr,\state_0}, \rew^\mathit{cr}, \timeb) .
  \intertext{Notice that, for $\sched \in \schedscr \cup \schedshr$, it is}
  \label{eqn:embschedeq}
    \emb(\cmodel_\sched) &= \emb(\cmodel)_\sched .
  \end{align}

  In addition,
  {\allowdisplaybreaks
  \begin{align}
      & \mvalue(\emb(\cmodel)_{\schedhr}, \state_0, \rew^{\mathit{hr}}, \timeb, \urate) \notag\\*
      = & \sum_{i=0}^\infty \left(\phi_{\urate \timeb}(i) \!\!\!\!\! \sum_{\state \in (\states\times\acts)^* \times \states} \!\!\!\!\! \tprob^{\emb(\cmodel)_{\schedhr}}(\state_0, i, \state) \cdot \frew^{\mathit{hr}}(\state)\right. \notag\\*
      & + \left.\psi_{\urate \timeb}(i) \!\!\!\!\! \sum_{\state \in (\states\times\acts)^* \times \states} \!\!\!\!\! \tprob^{\emb(\cmodel)_{\schedhr}}(\state_0, i, \state) \frac{\crew^{\mathit{hr}}(\state)}{\urate}\right) \notag\\
      = & \sum_{i=0}^\infty \left(\phi_{\urate \timeb}(i) \sum_{\state \in \states} \tprob^{\emb(\cmodel),{\schedhr}}(\state_0, i, \state) \cdot \frew(\state)\right. \notag\\*
      & + \left.\psi_{\urate \timeb}(i) \sum_{\state \in \states} \tprob^{\emb(\cmodel),{\schedhr}}(\state_0, i, \state) \frac{\crew(\state)}{\urate}\right) \label{eqn:mapscheds}\\
      \stackrel{\mbox{\tiny\refseqn{schedeqtrans}}}{=} &  \sum_{i=0}^\infty \left(\phi_{\urate \timeb}(i) \sum_{\state \in \states} \tprob^{\emb(\cmodel),{\schedcr}}(\state_0, i, \state) \cdot \frew(\state)\right. \notag\\*
      & + \left.\psi_{\urate \timeb}(i) \sum_{\state \in \states} \tprob^{\emb(\cmodel),{\schedcr}}(\state_0, i, \state) \frac{\crew(\state)}{\urate}\right) \notag\\
      = & \sum_{i=0}^\infty \left(\phi_{\urate \timeb}(i) \sum_{\state \in \states \times \nats} \tprob^{\emb(\cmodel)_{\schedcr}}((\state_0,0), i, \state) \cdot \frew^{\mathit{cr}}(\state)\right. \notag\\*
      & + \left.\psi_{\urate \timeb}(i) \sum_{\state \in \states \times \nats} \tprob^{\emb(\cmodel)_{\schedcr}}((\state_0,0), i, \state) \frac{\crew^{\mathit{cr}}(\state)}{\urate}\right) \notag\\
      = & \,\mvalue(\emb(\cmodel)_\schedcr, \state_0, \rew^{\mathit{cr}}, \timeb, \urate) .\notag
  \end{align}
  }
  
  From these facts, we have
  \begin{equation}
    \begin{split}
      & \mvalue(\stopro^{\cmodel,\schedhr,\state_0}, \rew, \timeb) \\
      & \stackrel{\mbox{\tiny\refseqn{hrsrewrewpr}}}{=} \mvalue(\stopro^{\cmodel_{\schedhr},\state_0}, \rew^{\mathit{hr}}, \timeb) \\
      & \stackrel{\mbox{\tiny\reflem{disc-rew}}}{=}
      \mvalue(\emb(\cmodel_{\schedhr}), \state_0, \rew^{\mathit{hr}}, \timeb, \urate) \\
      & \stackrel{\mbox{\tiny\refseqn{embschedeq}}}{=} \mvalue(\emb(\cmodel)_{\schedhr}, \state_0, \rew^{\mathit{hr}}, \timeb, \urate) \\
      & \stackrel{\mbox{\tiny\refseqn{mapscheds}}}{=} \mvalue(\emb(\cmodel)_\schedcr, \state_0, \rew^{\mathit{cr}}, \timeb, \urate) \\
      & \stackrel{\mbox{\tiny\refseqn{embschedeq}}}{=} \mvalue(\emb(\cmodel_\schedcr), \state_0, \rew^{\mathit{cr}}, \timeb, \urate) \\
      & \stackrel{\mbox{\tiny\reflem{disc-rew}}}{=} \mvalue(\stopro^{\cmodel_\schedcr,\state_0}, \rew^{\mathit{cr}}, \timeb) \\
      & \stackrel{\mbox{\tiny\refseqn{crrewrewpr}}}{=} \mvalue(\stopro^{\cmodel,\schedcr,\state_0}, \rew, \timeb) .
    \end{split}
  \end{equation}

  If we start with a CR $\sigma_\mathit{cr}'$, we can define the HR $\sigma_\mathit{hr}'$ such that,
  for $\mpath \in (\states \times \acts)^n$ with $n \in \nats$, we have
  $\sigma_\mathit{hr}'(\mpath, \state) \defeq \sigma_\mathit{cr}'(\state, n)$,
  and then the result can be shown in the same way.
\end{proof}

From \reflem{cr-sameval-hrs}, we can conclude that the maximum is obtained by a scheduler in $\schedscr$.
\begin{corollary}
  \label{cor:cr-suffices-ct}
  Given a CTMDP $\cmodel$, and reward structure $\rew$, for all $\state_0 \in \states$, we have
  \[
  \mvalue^{\max}(\cmodel, \state_0, \rew, \timeb) = \max_{\sched \in \schedscr} \mvalue(\stopro^{\cmodel,\sched,\state_0},\rew, \timeb) .
  \]
\end{corollary}

Because of \refcor{cr-suffices-ct}, we only have to show that \refalg{opt-reach} computes the maximum over all $\sched \in \schedscr$ up to the specified precision.

We first show that, to compute the value of a given CTMDP for a given scheduler up to a required precision, it suffices to consider a limited number of steps in the embedded model.
\begin{lemma}
  \label{lem:precbound:appendix}
  Given a CTMDP $\cmodel = (\states, \acts, \pmat)$ with a reward structure $\rew = (\crew, \frew)$, a precision $\precision > 0$, and $k$ such that
  \[
  \sum_{n=0}^k\psi_{\urate \timeb}(n) > \urate \timeb - \frac{\precision \urate}{2\crew^{\max}} \mbox{ and } \psi_{\urate \timeb}(k) \cdot \frew^{\max} < \frac{\precision}{2},
  \]
  then for all schedulers $\sched \in \schedscr$, and all $\state_0 \in \states$, we have
\begin{equation*}
\begin{split}
  & \sum_{i=k+1}^\infty \left(\phi_{\urate \timeb}(i) \sum_{\state \in \states} \tprob^{\emb(\cmodel),\sched}(\state_0, i, \state) \frew(\state)\right. \\
  & + \left.\psi_{\urate \timeb}(i) \sum_{\state \in \states} \tprob^{\emb(\cmodel),\sched}(\state_0, i, \state)
    \frac{\crew(\state)}{\urate}\right) < \precision .
\end{split}
\end{equation*}
\end{lemma}

\begin{proof}
Consider a CTMC $\cmodel' = (\states', \pmat')$ with reward structure $\rew' = (\crew', \frew')$, and
\begin{equation*}
\mvalue(\cmodel', \state_0, \rew', \timeb) = \underbrace{\expect \left[\int_0^\timeb \! \crew'(\stopro_u^{\cmodel',\state_0}) \, \diff u \right]}_{\mathrm{accumulated}} + \underbrace{\expect \left[\frew'(\stopro_\timeb^{\cmodel',\state_0}) \right]}_{\mathrm{final}}
\end{equation*}
for any $\state_0 \in \states$.
It is known~\cite[remark below Theorem 2]{KwiatkowskaNP06} that, if
\begin{equation*}
\sum_{n=0}^k\psi_{\urate \timeb}(n) > \urate \timeb - \frac{\precision \urate}{2\crew'^{\max}},
\end{equation*}
then
\begin{equation*}
\begin{split}
  & \mathrm{accumulated} \\
  \defeq {} & \sum_{n=k+1}^\infty \psi_{\urate \timeb}(i) \sum_{\state \in \states'} \tprob^{\emb(\cmodel')}(\state_0, i, \state) \frac{\crew'(\state)}{\urate} \\
  < {} & \frac{\precision}{2};
\end{split}
\end{equation*}
and if
\begin{equation*}
\psi_{\urate \timeb}(k) \cdot \frew'^{\max} < \frac{\precision}{2},
\end{equation*}
then
\begin{equation*}
\begin{split}
  & \mathrm{final} \\
  \defeq {} & \sum_{i=k+1}^\infty \phi_{\urate \timeb}(i) \sum_{\state \in \states'} \tprob^{\emb(\cmodel')}(\state_0, i, \state) \cdot \frew'(\state) \\
  \leq {} & \sum_{i=k+1}^\infty \phi_{\urate \timeb}(i) \sum_{\state \in \states'} \tprob^{\emb(\cmodel')}(\state_0, i, \state) \cdot \frew'^{\max} \\
  = {} & \sum_{i=k+1}^\infty \phi_{\urate \timeb}(i) \frew'^{\max} \\
  = {} & \psi_{\urate \timeb}(k) \cdot \frew'^{\max} \\
  < {} & \frac{\precision}{2} .
\end{split}
\end{equation*}

Thus,
\begin{equation}
\begin{split}
  & \sum_{i=k+1}^\infty \left(\phi_{\urate \timeb}(i) \sum_{\state \in \states} \tprob^{\emb(\cmodel')}(\state_0, i, \state) \frew'(\state)\right. \\
  & + \left.\psi_{\urate \timeb}(i) \sum_{\state \in \states} \tprob^{\emb(\cmodel')}(\state_0, i, \state)
    \frac{\crew'(\state)}{\urate}\right) \\
  = {} & \mathrm{accumulated} + \mathrm{final} \\
  < {} & \precision .
\end{split}
\end{equation}

Now, with $\rew' \defeq (\crew', \frew')$ where $\crew'(\state, n) \defeq \crew(\state)$, and $\frew'(\state, n) \defeq \frew(\state)$,
because $\crew^{\max} = \crew'^{\max}$ and $\frew^{\max} = \frew'^{\max}$ the following holds.
\begin{equation*}
\begin{split}
  & \sum_{i=k+1}^\infty \left(\phi_{\urate \timeb}(i) \sum_{\state \in \states} \tprob^{\emb(\cmodel),\sched}(\state_0,
  i, \state) \cdot \frew(\state)\right. \\
  & + \left.\psi_{\urate \timeb}(i) \sum_{\state \in \states} \tprob^{\emb(\cmodel),\sched}(\state_0, i, \state)
    \frac{\crew(\state)}{\urate}\right) 
\end{split}
\end{equation*}
\begin{equation*}
\begin{split}
  = & \sum_{i=k+1}^\infty \left(\phi_{\urate \timeb}(i) \sum_{\state \in \states \times \nats} \tprob^{\emb(\cmodel)_\sched}(\state_0,
  i, \state) \cdot \frew'(\state)\right. \\
  & + \left.\psi_{\urate \timeb}(i) \sum_{\state \in \states \times \nats} \tprob^{\emb(\cmodel)_\sched}(\state_0, i, \state)
    \frac{\crew'(\state)}{\urate}\right) \\
  < {} & \precision .
\end{split}
\end{equation*}
\end{proof}

\begin{algorithm}
  \caption{\label{alg:sched-value} Compute value of $\cmodel = (\states, \acts, \rmat)$, reward structure $(\crew, \frew)$ and CR $\sched$ up to precision $\precision$.}
  \DontPrintSemicolon
  let $k$ s.t. $\sum_{n=0}^k\psi_{\urate \timeb}(n) > \urate \timeb -
  \precision \urate/(2\crew^{\max})
  \wedge \psi_{\urate \timeb}(k) \cdot \frew^{\max} <
  \precision/2 $\;
  $\cmodel' = (\states, \acts, \pmat) \gets \emb(\cmodel)$ \;
  \lForAll{$\state \in \states$} {%
    $q_{k+1}(\state) \gets 0$ \;
  }
  \ForAll{$i=k,k-1,\ldots,0$} {
    \ForAll{$\state \in \states$} {
      $m \gets \sum_{\act \in \acts} \sched(\state,k)(\act) \sum_{\state' \in S} \pmat(s,\act,s')
      q_{i+1}(s')$ \;
      $q_i(\state) \gets m + \phi_{\urate \timeb}(i) \cdot \frew(\state) +
      \psi_{\urate \timeb}(i) \cdot \crew(\state)/\urate$ \;
    }
  }
  \Return $q_0$
\end{algorithm}

We can show that \refalg{sched-value} computes the values of a CTMDP given a certain scheduler up to a required precision.
\begin{lemma}
\label{lem:sched-value-prec}
Consider a CTMDP $\cmodel = (\states, \acts, \rmat)$, a time bound $\timeb$, a scheduler $\sched \in \schedscr$, and let $q$ be the return value of \refalg{sched-value}.
Then $|q(\state_0) - \mvalue(\stopro^{\cmodel,\sched,\state_0}, \rew, \timeb) | < \precision$ for all $\state_0 \in \states$.
\end{lemma}

\begin{proof}
Consider the values $q = q_0$ returned by \refalg{sched-value}.
It holds that
\begin{equation}
\begin{split}
  & q(s_0) \\
  = {} & \phi_{\urate \timeb}(0) \cdot \frew(\state_0) + \psi_{\urate \timeb}(0) \frac{\crew(\state_0)}{\urate} \\
  & + \sum_{\act_0 \in \acts} \sched(\state_0,0)(\act_0) \sum_{\state_1 \in \states} \pmat(\state_0, \act_0, \state_1) \\
  & ~~~~~~~~~~\cdot(\phi_{\urate \timeb}(1) \cdot \frew(\state_1) + \psi_{\urate \timeb}(1) \frac{\crew(\state_1)}{\urate} \\
  & ~~~~~~~~~~~~~~~ +\!\! \sum_{\act_1 \in \acts} \sched(\state_1,1)(\act_1)\! \sum_{\state_2 \in \states} \pmat(\state_1, \act_1, \state_2)\cdots \\
  = {} &  \phi_{\urate \timeb}(0) \cdot \frew(\state_0) + \psi_{\urate \timeb}(0) \frac{\crew(\state_0)}{\urate} \\
  & + \sum_{\act_0 \in \acts }\sched(\state_0,0)(\act_0) \sum_{\state_1 \in \states} \pmat(\state_0, \act_0, \state_1)( \phi_{\urate \timeb}(1) \cdot \frew(\state_0) \\
  & ~~~~~~~~~~~~~~~~~~~~~~~~~~~~~~~~~~~~+ \psi_{\urate \timeb}(1) \frac{\crew(\state_0)}{\urate}) \\
  & + \!\!\sum_{\act_0 \in \acts}\sched(\state_0,0)(\act_0)\!\! \sum_{\state_1 \in \states} \pmat(\state_0, \act_0, \state_1) \!\!\sum_{\act_1 \in \acts} \!\! \sched(\state_1, 1)\cdots \\
  & + \cdots \\
  \stackrel{\mbox{\tiny\refscor{tprob-matrix}}}{=} & \sum_{i=0}^k \left(\phi_{\urate \timeb}(i) \sum_{\state \in \states} \tprob^{\emb(\cmodel),\sched}(\state_0,
  i, \state) \frew(\state)\right. \\
  & + \left.\psi_{\urate \timeb}(i) \sum_{\state \in \states} \tprob^{\emb(\cmodel),\sched}(\state_0, i, \state)
    \frac{\crew(\state)}{\urate}\right) \\
  = {} &  \mvalue(\emb(\cmodel), \state_0, \rew, \timeb, \urate) - \\
  & \sum_{i=k+1}^\infty \left(\phi_{\urate \timeb}(i) \sum_{\state \in \states} \tprob^{\emb(\cmodel),\sched}(\state_0,
  i, \state) \frew(\state)\right. \\
  & + \left.\psi_{\urate \timeb}(i) \sum_{\state \in \states} \tprob^{\emb(\cmodel),\sched}(\state_0, i, \state)
    \frac{\crew(\state)}{\urate}\right) \\
  \stackrel{\mbox{\tiny\reflem{precbound}}}{=} &  \mvalue(\emb(\cmodel), \state_0, \rew, \timeb, \urate) - \precision'
\end{split}
\end{equation}
for some $\precision'$ with $0 \leq \precision' < \precision$, and thus $|q(\state_0) - \mvalue(\stopro^{\cmodel,\sched,\state_0}, \rew, \timeb) | < \precision$.
\end{proof}

The value obtained from \refalg{opt-reach} will be no smaller than the one obtained from applying \refalg{sched-value} on an arbitrary scheduler.
\begin{lemma}
\label{lem:opt-val-nosmaller}
Consider a CTMDP $\cmodel = (\states, \acts, \rmat)$, a time bound $\timeb$, and an arbitrary scheduler $\sched \in \schedscr$;
let $q$ be the return value of \refalg{sched-value}, and let $q'$ be the return value of \refalg{opt-reach}.
Then $q(\state_0) \leq q'(\state_0)$ holds for all $\state_0 \in \states$.
\end{lemma}

\begin{proof}
  Let $q_i$ be as given in \refalg{sched-value}, and let $q_i'$ be the corresponding vector of \refalg{opt-reach}.
  We show the lemma by backward induction on the program variable $i$.

  Induction start: $i = k+1$. Before the main loop at the lines 3, both algorithms assign $q_{k+1}(\state) = q_{k+1}'(\state) = 0$ for all $\state \in \states$.

  Induction assumption: Assume it is $q_{i+1}(\state) \leq q'_{i+1}(\state)$ at the beginning of the main loops, that is before line~4.

  Induction step: Consider $m$ of \refalg{sched-value}, and corresponding $m'$ of \refalg{opt-reach} after the assignment to this variable at line~6.
  We have
  \begin{equation*}
  \begin{split}
    m & = \sum_{\act \in \acts(\state)} \sched(\state,k)(\act) \sum_{\state' \in S} \pmat(s,\act,s')
    q_{i+1}(s') \\
    & \leq \max_{\act \in \acts(\state)} \sum_{\state' \in S} \pmat(s,\act,s')
    q_{i+1}(s') \\
    & \stackrel{\mbox{\tiny Ass.}}{\leq} \max_{\act \in \acts(\state)} \sum_{\state' \in S} \pmat(s,\act,s')
    q_{i+1}'(s') \\
    & = m' .
  \end{split}
  \end{equation*}
  Because lines 7 are identical in both algorithms, also $q_i \leq q_i'$ at the end of the main loops.  
\end{proof}

With these preparations, we can now prove the first part of~\refpro{algo-correctness}.
\begin{lemma}
\label{lem:algo-correctness}
Consider a CTMDP $\cmodel = (\states, \acts, \rmat)$, a reward structure $\rew$, a time bound $\timeb$, and let $q'$ be the return value of \refalg{opt-reach}.
Then, for all $\state_0 \in \states$, it is $|q'(\state_0) - \mvalue^{\max}(\cmodel, \state_0, \rew, \timeb)| < \precision$.
\end{lemma}

\begin{proof}
  Let $\sched \in \schedscr$ be such that
  \begin{equation}\label{eqn:opt-schedscr}
    \mvalue(\stopro^{\cmodel,\sched,\state_0}, \rew, \timeb) = \mvalue^{\max}(\cmodel, \state_0, \rew, \timeb).
  \end{equation}
  Then, because of \reflem{sched-value-prec}, $|q(\state_0) - \mvalue(\stopro^{\cmodel,\sched,\state_0}, \rew, \timeb) | < \precision$ is true,
  where $q$ is the return value of \refalg{sched-value}.
  Because of \reflem{opt-val-nosmaller}, we know that $q(\state_0) \leq q'(\state_0)$.
  By adding the assignment
  \begin{equation*}
    \sigma_\mathit{cd}'(\state, i) \gets \arg\max_{\act\in\acts(s)} \sum_{\state' \in S} \pmat(s,\act,s') q_{i+1}(s')
  \end{equation*}
  into the inner loop of \refalg{opt-reach} after \reflin{last-inner}, we can obtain a prefix of the scheduler $\sigma_\mathit{cd}' \in \schedscd$.
  Consider $\sigma_\mathit{cr}' \in \schedscr$ such that $\sigma_\mathit{cr}'(\state, i)(\act) = 1$ if $\sigma_\mathit{cd}'(\state, i) = \act$,
  and $\sigma_\mathit{cr}'(\state, i)(\act) = 0$ else.
  It can easily be shown that applying \refalg{sched-value} on $\sigma_\mathit{cr}'$ also yields the value $q'$.
  Thus, again by \reflem{sched-value-prec}, we have $|q'(\state_0) - \mvalue(\stopro^{\cmodel,\sched,\state_0}, \rew, \timeb) | < \precision$.
\end{proof}

We can now show that deterministic schedulers suffice, by using the fact that \refalg{opt-reach} indeed maximises only over this class.
\begin{lemma}
  Let $\cmodel = (\states, \acts, \rmat)$ be a CTMDP with reward structure $\rew = (\crew, \frew)$.
  Then there exists $\sched \in \schedscd$ such that $\mvalue^{\max}(\cmodel, \state_0, \rew, \timeb) = \mvalue(\stopro^{\cmodel,\sched,\state_0},\rew, \timeb)$.
\end{lemma}

\begin{proof}
  Assume the lemma does not hold.
  Then, for each $\sched \in \schedscd$, there is a $\precision > 0$ such that, for some state $\state_0 \in \states$,
  we have $|\mvalue(\stopro^{\cmodel,\sched,\state_0}, \rew, \timeb) - \mvalue^{\max}(\cmodel, \state_0, \rew, \timeb)| \geq \precision$.
  Now consider $\sched_\mathit{cd}' \in \schedscd$ obtained by \refalg{opt-reach} as in the proof of \reflem{algo-correctness} using the same required precision $\precision$.
  By the correctness of \refalg{opt-reach}, we then have $|\mvalue(\stopro^{\cmodel,\sched',\state_0}, \rew, \timeb) - \mvalue^{\max}(\cmodel, \state_0, \rew, \timeb)| < \precision$.
  This result contradicts the assumption, because with the extension in the proof of \reflem{algo-correctness},
  the algorithm also computes a CD which obtains this precision.
\end{proof}
The idea of using the fact that the optimising algorithm computes a scheduler of a more restricted class than the class considered was adapted from various proofs for similar problems in the discrete-time setting~\cite{Puterman94}.

\subsection*{Proof of \refpro{corr-abs}}
\label{apx:corr-abs}
\noindent
Consider $(\states, \pmat) = \dmodel \defeq \emb(\cmodel)$, and
$(\states, \pmat') = \dmodel' \defeq \emb(\cmodel')$.
We define the HR
$\sched\colon \bigl((\astates \times \acts)^* \times \astates\bigr) \to \distrs(\acts)$
so that for $\mpath = \astate_0 \act_{\state_0} \astate_1 \act_{\state_1} \ldots \act_{\state_{n-1}} \astate_n$ we have
\[
\sched(\mpath)(\act_\state) \defeq \prob(\stopro^{\dmodel,\state_0}_n = \state \mid \stopro^{\dmodel,\state_0}_n \wedge \bigwedge_{i=0}^{n-1} \stopro^{\dmodel,\state_0}_i = \state_i) .
\]
By induction, for all $n \in \nats$, and $\astate \in \astates$, we have
\[
\prob(\stopro^{\dmodel,\state_0}_n \in \astate) = \prob(\stopro^{\dmodel',\sched,\astate_0}_n = \astate) .
\]
Thus, using \refdef{disc-value}, \refdef{induced}, and the definition of the reward structures $\rew,\rew'$ it is
\[
\mvalue(\dmodel, \state_0, \rew, \timeb, \urate)
\leq \mvalue(\dmodel_\sched', \astate_0, \rew, \timeb, \urate);
\]
and thus using \reflem{disc-rew} we obtain
\[
\mvalue(\stopro^{\cmodel,\state_0}, \rew, \timeb)
\leq \mvalue(\stopro^{\cmodel',\sched,\astate_0},\rew',\timeb)
\leq \mvalue^{\max}(\cmodel', \astate_0, \rew', \timeb) .
\]

\subsection*{Proof of \refpro{monotone}}
\label{apx:monotone}
\noindent
We only show that, by using a finer abstraction, the maximal bound cannot increase.
The case for the minimal bound is likewise.
We have to show
\begin{equation}
  \label{eqn:rfine-eps-orig}
  \mvalue^{\max}(\cmodel, \astate_t, \rew, \timeb) \geq \mvalue^{\max}(\cmodel', \astate_{t,j}', \rew', \timeb) .
\end{equation}

We will first show that, for each $\precision > 0$, it holds that
\begin{equation}
  \label{eqn:rfine-eps}
  \mvalue^{\max}(\cmodel, \astate_t, \rew, \timeb) + \precision \geq \mvalue^{\max}(\cmodel', \astate_{t,j}', \rew', \timeb) .
\end{equation}
To do so, we show by a backward induction on \refalg{opt-reach} that, for the value $q_0$ obtained for $\cmodel$, and the value $q'$ obtained for $\cmodel'$, we have $q \geq q_0'$, using a precision of $\precision$.
By the precision guarantee of the algorithm from \refpro{algo-correctness}, doing so in turn shows \refeqn{rfine-eps} for the given $\precision$.

Induction start with $i = k+1$. Before the main loop, at \reflin{init}, both runs assign $q_{k+1}(\astate_t) = q_{k+1}'(\astate_{t,j}') = 0$ for all $\astate_t \in \apart, \astate_{t,j}' \in \apart'$.

Induction assumption. Assume $q_{i+1}(\astate_t) \geq q'_{i+1}(\astate_{t,j}')$ holds
for all $\astate_t \in \apart, \astate_{t,j}' \in \apart'$ with $\astate_t = \bigcup_{j=1}^u \astate_{t,j}'$
at the beginning of the main loops, before \reflin{main-start}.

Induction step. Consider $m$ for $\astate_t$ of $\cmodel$, and corresponding $m'$ for $\astate_{t,j}'$ of $\cmodel'$, after the assignment to this variable at \reflin{asg-max}.
Let $(\hat{\act}', \act')$ be a maximising decision for some $\astate_{t,j}'$ at this line.
By the definition of the abstraction, for $\astate_t$, we find a corresponding $(\hat{\act}, \act)$ such that
\begin{itemize}
\item $\dom(\hat{\act}') = \dom(\hat{\act})$, and
\item for each $\prismCmd \in \dom(\hat{\act}')$ we have $\hat{\act}'(\prismCmd) = (\astate'_{v,l}, \mathbf{I})$ and $\hat{\act}(\prismCmd) = (\astate_{v}, \mathbf{I})$ with $\astate'_{v,l} \subseteq \astate_{v}$.
\end{itemize}
Then we have
\begin{equation*}
  \begin{split}
    m' & = \max_{(\hat{\act}', \act')\in\acts(\astate_{t,j}')} \sum_{\astate'_{v,l} \in \apart'} \pmat(\astate_{t,j},(\hat{\act}', \act'),\astate'_{v,l}) q_{i+1}(\astate'_{v,l}) \\
    & = \sum_{\astate'_{v,l} \in \apart'} \pmat(\astate_{t,j}',(\hat{\act}', \act'),\astate'_{v,l}) q_{i+1}(\astate'_{v,l}) \\
    & \stackrel{\mbox{\tiny Ass.}}{\leq} \sum_{\astate_v \in \apart} \pmat(\astate_t,(\hat{\act}, \act),\astate_v) q_{i+1}(\astate_v) \\
    & \leq \max_{(\hat{\act}, \act)\in\acts(\astate_t)} \sum_{\astate_v \in \apart} \pmat(\astate_t,(\hat{\act}, \act),\astate_v) q_{i+1}(\astate_v) \\
    & = m .
  \end{split}
\end{equation*}
The rewards for the refined abstraction cannot be larger than the ones for the coarser one.
Thus, after \reflin{last-inner}, we still have $q_i \geq q_i'$ at the end of each iteration.

The validity of \refeqn{rfine-eps} then also shows that the inequality \refeqn{rfine-eps-orig} holds for $\precision = 0$:
if $\mvalue^{\max}(\cmodel, \astate_t, \rew, \timeb) < \mvalue^{\max}(\cmodel', \astate_{t,j}', \rew', \timeb)$, then $\precision' \defeq \mvalue^{\max}(\cmodel', \astate_{t,j}', \rew', \timeb) - \mvalue^{\max}(\cmodel, \astate_t, \rew, \timeb)$ is positive.
By subtracting $\mvalue^{\max}(\cmodel, \astate_t, \rew, \timeb)$ from \refeqn{rfine-eps}, we have
\begin{equation*}
  \precision \geq \precision' .
\end{equation*}
This equation must hold for all $\precision$, for instance $\precision'/2$.
Thus, we would obtain a contradiction if it would not hold for $\precision = 0$.

\qed

\balance

\bibliographystyle{plain}
\bibliography{bib}

\clearpage

\textit{Ernst Moritz Hahn} received his Ph.D.\ from Saarland University in 2013.
He wrote his
doctoral thesis at the chair of Dependable Systems and Software, advised
by Prof. Dr.-Ing. Holger Hermanns. After his dissertation, he became
research assistant at the group of automated verification at the
University of Oxford in the VERIWARE project lead by Marta Kwiatkowska.
Ernst Moritz Hahn is currently associate professor in the Institute of
Software Chinese Academy of Sciences. His research area is in
probabilistic verification, such as in the analysis of very large Markov chains,
probabilistic hybrid systems, and parametric Markov models.

\medskip

\textit{Holger Hermanns} is a full professor at the Department of Computer Science at Saarland University, Saarbr{\"u}cken, Germany, holding the chair for Dependable Systems and Software.
His research interests include compositional modeling and verification of concurrent systems, resource-aware embedded systems, and performance and dependability evaluation of critical infrastructure.
In these areas, Holger Hermanns has authored or co-authored more than 150 peer-reviewed scientific papers (ha-index 92).
He serves on the steering committees of ETAPS, TACAS, and QEST.
He is Member of Academia Europaea. 

\medskip

\textit{Ralf Wimmer} received his diploma with distinction 
in computer science from the Albert-Ludwigs-Universit\"at Freiburg, 
Germany in 2004. Afterwards, he
worked as a Ph.D.\ student at the Chair of Computer Architecture at the
same university, advised by Prof.\ Dr.\ Bernd Becker. He obtained his Ph.D.\ degree 
with distinction in 2011 for his thesis on symbolic methods for
probabilistic verification. Since then, he is continuing his work as a
research assistant and leader of the verification group at the Chair
of Computer Architecture. His research focus is on symbolic methods
and solver technologies, and their application for the verification of
digital and stochastic systems.

\medskip

\textit{Bernd Becker} is a Full Professor (Chair of Computer Architecture) at the 
Faculty of Engineering, University of Freiburg, Germany. Prior to joining the 
University of Freiburg in 1995, he was with J.~W.~Goethe-University Frankfurt as 
an associate professor for complexity theory and efficient algorithms. His 
research activities include design, test, and verification methods for embedded 
systems and  nanoelectronic circuitry. He is a Co-Speaker of the DFG 
Transregional Collaborative Research Center ``Automatic Analysis and Verification 
of Complex Systems (AVACS),'' and a Director of the Centre for Security and 
Society, University of Freiburg. He is a fellow of IEEE, and Member of Academia 
Europaea.

\balance

\end{document}